\documentclass[useAMS,usenatbib,usegraphicx]{mn2e}
\usepackage[total={17.8cm,24.0cm}, centering]{geometry}
\usepackage{times}

\usepackage{graphicx}
\usepackage{epsfig}
\usepackage{amssymb}
\usepackage{natbib}
\usepackage{journals}
\usepackage{deluxetable}
\usepackage[latin1]{inputenc}
\newcommand{\Sauron}{\texttt{SAURON}}
\newcommand{\atlas}{{ATLAS$^{\rm 3D}$}}
\newcommand{\XSauron}{\texttt{XSauron}}

\newcommand{\Msun} {$\mbox{M}_{\sun}$}

\newcommand{\lr}{\,{$\lambda_R$}}

\def\spose#1{\hbox to 0pt{#1\hss}}
\def\lta{\mathrel{\spose{\lower 3pt\hbox{$\sim$}}
    \raise 2.0pt\hbox{$<$}}}
\def\gta{\mathrel{\spose{\lower 3pt\hbox{$\sim$}}
    \raise 2.0pt\hbox{$>$}}}
\newdimen\hssize
\newdimen\hdsize
\title[The \atlas\ project -- III. A census of the stellar angular momentum in ETGs]
 {The \atlas\ project -- III. A census of the stellar angular momentum within the
 effective radius of early-type galaxies: unveiling the distribution of Fast  
 and Slow Rotators}
\author
[Eric Emsellem et al.]{\parbox{\textwidth}{Eric Emsellem,$^{1,2}$\thanks{E-mail: eric.emsellem@eso.org \texttt{}}
Michele Cappellari,$^{3}$
Davor Krajnovi\'c,$^{1}$
Katherine Alatalo,$^{4}$
Leo Blitz,$^{4}$
Maxime Bois,$^{1,2}$
Fr\'ed\'eric Bournaud,$^{5}$
Martin Bureau,$^{3}$
Roger L. Davies,$^{3}$
Timothy A. Davis,$^{3}$
P. T. de Zeeuw,$^{1,6}$
Sadegh Khochfar,$^{7}$
Harald Kuntschner,$^{8}$
Pierre-Yves Lablanche,$^{2}$
Richard M. McDermid,$^{9}$
Raffaella Morganti,$^{10,11}$
Thorsten Naab,$^{12}$
Tom Oosterloo,$^{10,11}$
Marc Sarzi,$^{13}$
Nicholas Scott,$^{3}$
Paolo Serra,$^{10}$
Glenn van de Ven,$^{14}$
Anne-Marie Weijmans,$^{15}$\thanks{Dunlap Fellow}
and Lisa M. Young,$^{16}$}\vspace{0.4cm}\\ 
\parbox{\textwidth}{$^{1}$European Southern Observatory, Karl-Schwarzschild-Str. 2, 85748 Garching, Germany\\
$^{2}$Universit\'e Lyon 1, Observatoire de Lyon, Centre de Recherche Astrophysique de Lyon \\ \hspace*{0.5cm} and Ecole Normale Sup\'erieure de Lyon, 9 avenue Charles Andr\'e, F-69230 Saint-Genis Laval, France\\
$^{3}$Sub-department of Astrophysics, Department of Physics, University of Oxford, Denys Wilkinson Building, Keble Road, Oxford OX1 3RH, UK\\
$^{4}$Department of Astronomy, Campbell Hall, University of California, Berkeley, CA 94720, USA\\
$^{5}$Laboratoire AIM Paris-Saclay, CEA/IRFU/SAp -- CNRS -- Universit\'e Paris Diderot, 91191 Gif-sur-Yvette Cedex, France\\
$^{6}$Sterrewacht Leiden, Leiden University, Postbus 9513, 2300 RA Leiden, the Netherlands\\
$^{7}$Max-Planck Institut f\"ur extraterrestrische Physik, PO Box 1312, D-85478 Garching, Germany\\
$^{8}$Space Telescope European Coordinating Facility, European Southern Observatory, Karl-Schwarzschild-Str. 2, 85748 Garching, Germany\\
$^{9}$Gemini Observatory, Northern Operations Centre, 670 N. A`ohoku Place, Hilo, HI 96720, USA\\
$^{10}$Netherlands Institute for Radio Astronomy (ASTRON), Postbus 2, 7990 AA Dwingeloo, The Netherlands\\
$^{11}$Kapteyn Astronomical Institute, University of Groningen, Postbus 800, 9700 AV Groningen, The Netherlands\\
$^{12}$Max-Planck-Institut f\"ur Astrophysik, Karl-Schwarzschild-Str. 1, 85741 Garching, Germany\\
$^{13}$Centre for Astrophysics Research, University of Hertfordshire, Hatfield, Herts AL1 9AB, UK\\
$^{14}$Max-Planck-Institut f\"ur Astronomie, K\"onigstuhl 17, 69117 Heidelberg, Germany\\
$^{15}$Dunlap Institute for Astronomy \& Astrophysics, University of Toronto, 50 St. George Street, Toronto, ON M5S 3H4, Canada\\
$^{16}$Physics Department, New Mexico Institute of Mining and Technology, Socorro, NM 87801, USA
}}
\begin{document}
\maketitle
%
%
\clearpage
\begin{abstract}
We provide a census of the apparent stellar angular momentum within one effective radius of a volume-limited sample of 260 early-type galaxies (ETGs) in the nearby Universe, using integral-field spectroscopy obtained in the course of the \atlas\ project.
We exploit the $\lambda_R$ parameter (previously used via a constant threshold value of 0.1) to characterise the existence of two families of ETGs: Slow Rotators which exhibit complex stellar velocity fields and often include stellar kinematically distinct cores, and Fast Rotators which have regular velocity fields. Our complete sample of 260 ETGs leads to a new criterion to disentangle Fast and Slow Rotators which now includes a dependency on the apparent ellipticity $\epsilon$. It separates the two classes significantly better than the previous prescription, and than a criterion based on $V/\sigma$: Slow Rotators and Fast Rotators have ${\lambda_R}$ lower and larger than $k_{FS} \times \sqrt{\epsilon}$, respectively, where $k_{FS} = 0.31$ for measurements made within an effective radius $R_e$.

We show that the vast majority of early-type galaxies are Fast Rotators: these have regular stellar rotation, with aligned photometric and kinematic axes \citep[][Paper~II]{Krajnovic+11}, include discs and often bars and represent $86\pm2$\% (224/260) of all early-type galaxies in the volume-limited \atlas\ sample. Fast Rotators span the full range of apparent ellipticities from $\epsilon = 0$ to 0.85, and we suggest that they cover intrinsic ellipticities from about 0.35 to 0.85, the most flattened having morphologies consistent with spiral galaxies. Only a small fraction of ETGs are Slow Rotators representing $14\pm2$\% (36/260) of the \atlas\ sample of ETGs. Of all Slow Rotators, 11\% (4/36) exhibit two counter-rotating stellar disc-like components and are rather low mass objects ($M_{\rm dyn} < 10^{10.5}$~M$_{\odot}$). All other Slow Rotators (32/36) appear relatively round on the sky ($\epsilon_e < 0.4$), tend to be massive ($M_{\rm dyn} > 10^{10.5}$~M$_{\odot}$), and often (17/32) exhibit Kinematically Distinct Cores. Slow Rotators dominate the high mass end of ETGs in the \atlas\ sample, with only about one fourth of galaxies with masses above $10^{11.5}$~M$_{\odot}$ being Fast Rotators.
We show that the $a_4$ parameter which quantifies the isophotes disciness or boxiness does not seem to be simply related with the observed kinematics, while our new criterion based on $\lambda_R$ and $\epsilon$ is nearly independent from the viewing angles.
We further demonstrate that the separation of ETGs in E's (ellipticals) and S0's (lenticulars) is misleading. Slow and Fast Rotators tend to be classified as ellipticals and lenticulars, respectively, but the contamination is strong enough to affect results solely based on such a scheme: 20\% of all Fast Rotators are classified as E's, and more importantly 66\% of all E's in the \atlas\ sample are Fast Rotators.

Fast and Slow Rotators illustrate the variety of complex processes shaping galactic systems, such as e.g., secular evolution, disc instabilities, interaction and merging, gas accretion, stripping and harassment, forming a sequence from high to low (stellar) baryonic angular momentum. Massive Slow Rotators represent the extreme instances within the red sequence of galaxies which might have suffered from significant merging without being able to rebuild a fast rotating component within one effective radius. We therefore argue for a shift in the paradigm for early-type galaxies, where the vast majority of ETGs are galaxies consistent with nearly oblate systems (with or without bars), and where only a small fraction of them (less than 12\%) have central (mildly) triaxial structures.
\end{abstract}

\section{Introduction\label{sec:intro}}

Early-type galaxies lie at one end of the Hubble tuning fork, the schematic ordering of
galaxy morphology classes established long ago \citep{Hubble36, RC3}, the late-type
spirals being located at the other end of the diagram (excluding irregulars).
Although the Hubble sequence is not thought to represent an evolutionary sequence,
early-type systems are generally considered to be the output of
violent and extreme processes mainly driven by interactions and mergers in a
hierarchical Universe, constrasting with formation processes advocated for
spiral galaxies.

At moderate to high redshifts, the difficulty to obtain detailed photometric and spectroscopic 
information of large numbers of systems often lead to gather early-type galaxies into a single class: 
early-type galaxies are generally viewed as a single family of objects,
separated from the line of spiral systems, and which can
be studied at various redshifts \citep[e.g.][]{Bell+04, McIntosh+05, Kriek+08}.
There are nowadays various techniques to build samples of early-type
galaxies from large surveys, e.g., colour/spectroscospic information based on 
the fact that most galaxies in the red sequence are early-type 
\cite[see e.g.][and references therein]{Bell+04}. This obviously includes visual 
classification which, when applied at relatively lower redshift,
can lead to a classification closer to the full-fledged schemes defined by Hubble or de
Vaucouleurs \citep{Fukugita+07}. 

For nearby samples of galaxies, early-type galaxies are more commonly separated into 
two groups, namely the ellipticals (E's) and lenticulars (S0's), 
the latter being a transition class to (or from) the spiral systems \citep{Hubble36}. 
This is the existing {\em paradigm for early-type galaxies}, nicely illustrated 
by e.g., the work conducted by \cite{Bernardi+10} who studied a large set of galaxies extracted
from the Sloan Digital Sky Survey. 
Early-type galaxies are mostly red-sequence objects, comprising
E's (ellipticals) and S0's (lenticulars), each class representing about {\em half of
such a magnitude limited sample of early-type galaxies}. All together, Es and S0s
represent a significant fraction ($\sim 40\%$) of the total stellar mass density in the
nearby Universe \citep{Bernardi+10,Fukugita+07}. 
Ellipticals are thought to be nearly pure spheroidal objects, 
sometimes with central discs, while lenticulars are disc galaxies 
with large bulges/spheroidal components. Ellipticals are on average rounder 
than lenticulars, with very few E's having ellipticities higher than 0.4, 
and dominate the high-mass range of early-type galaxies.

This picture has been regularly and significantly updated 
specifically for galaxies in the nearby Universe for which we often have exquisite
photometric and spectroscopic information \citep{Bender+94,Rix+99, Gerhard+99}. 
An attempt to, for instance, further refine the elliptical E 
class into the boxy and discy systems was pursued by \cite{KB96}:
the proposed classification recognises the sequence of
intrinsic flattening and shapes and tries to address the presence of discs 
\citep{RixWhite90,Scorza+98, NaabBurkert01} 
and the dynamical status of these galaxies via a photometric proxy 
(namely, $a_4$ representing part of the
deviation of isophotes from pure ellipses). This has been recently expanded in
the context of "light deficit/excess" \citep{Graham+03,Graham04, Ferrarese+06, Cote+06,
Kormendy+09} to examine whether or not different groups of ellipticals may 
be key to link photometric properties with
their formation and assembly scenarios \citep{Naab+99, Khochfar+05, Kormendy+09,Hopkins+09a, Hopkins+09c}.
Nevertheless, early-type galaxies continue to be divided into ellipticals (spheroidal-like) and
lenticular (disc-like) systems, the former exhibiting some mild triaxiality
(usually associated with anisotropy), while the latter are prone to 
typical disc perturbations, such as e.g., bars.

There are many complications associated to these classification schemes due to e.g., inclination
effects or the limitations of photometric measurements \citep{KB96,PaperX}.
It is hard to disentangle lenticulars from ellipticals, which limits
the conclusions from studies using these as prime ingredients.
\citet[][hereafter E+07]{PaperIX} have emphasised the fact that
stellar kinematics contain critical information on the actual dynamical status of
the galaxy. E+07 suggested that $\lambda_R$, a simple parameter which can be derived
from the first two stellar velocity moments, can be used as a robust estimator of the
apparent specific angular moment (in stars) of galaxies \citep[see also][]{Jesseit+09}. 

Following this prescription, E+07 and \citet[][hereafter C+07]{PaperX} have shown that early-type galaxies are
distributed within two broad families depending on their $\lambda_R$ values: Slow Rotators ($\lambda_R < 0.1$),
which show little or no rotation, significant misalignments between the
photometric and the kinemetric axes, and contain kinematically decoupled components;
and Fast Rotators ($\lambda_R \geq 0.1$) which exibit regular stellar velocity
fields, consistent with disc-like rotation \citep{PaperXII} and sometimes bars.
These results were, however, based on a representative (but not
complete) sample of 48 early-type galaxies that is not a fair sample of the
local Universe.

We are now in a position to re-examine these issues in the light of 
the volume-limited \atlas\ survey \citep[][hereafter, Paper~I]{Cappellari+11a}. 
In the present paper, we wish to establish a census of the
stellar angular momentum of early-type galaxies (ETGs hereafter) in their
central regions via $\lambda_R$, and examine how we can relate such a measurement to their formation
processes. We will show that a kinematic classification based on \lr\ is a more
natural and physically motivated way of classifying galaxies. 
More importantly, we find that galaxies classified as Fast Rotators,
based on a refined criterion for \lr, represent 86\% of this magnitude limited
sample of ETGs, spread over a large range of flattening, the higher end being
within the range covered by spirals. Slow Rotators comprise 14\% of the \atlas\ sample, with about 12\%
of massive early-type galaxies with very low rotation and often with
Kinematically Decoupled Components, plus about 2\% of lower-mass flattened systems 
which represent the special case of counter-rotating disc-like components.

In Section~\ref{sec:obs} we briefly describe the observations we are using for
this study. Section~\ref{sec:LambdaR} is dedicated to a first presentation of the
measurements, mainly $\lambda_R$ and its relation to other basic properties such
as ellipticity and dynamical mass. In Section~\ref{sec:slowfast} we use these measurements to
update our view on Fast and Slow Rotators and propose a new criterion based on
simple dynamical arguments. We discuss the corresponding results in
Section~\ref{sec:discussion} and conclude in Section~\ref{sec:conclusions}.


\section{Observational material and analysis} 
\label{sec:obs}

The \atlas\ project is based on a volume-limited sample of 260 targets
extracted from a complete sample of early-type galaxies (ETGs). A detailed
description of the selection process and properties for the sample are given
in \cite[][hereafter Paper~I]{Cappellari+11a}, so we only provide a summary here. The parent sample is
comprised of all galaxies within a volume of 42~Mpc, brighter than $M_K =
-21.5$~mag \citep[2MASS, see][]{Jarrett+00} and 
constrained by observability ($\left|{\delta - 29\degr}\right| < 35$
and away from the Galactic equatorial plane). All 871 galaxies were examined via DSS
and SDSS colour images to classify them as ETGs (e.g., absence of clear spiral arms; see Paper~I).

The present study mostly relies on integral-field spectroscopic data from the
\Sauron\ instrument \citep{Bacon+01} mounted on the William Herschel Telescope (La Palma, Canary Islands).
We have also made use of imaging data obtained from several facilities,
including SDSS DR7 \citep{SDSS+09}, INT and MDM (see Paper~I for details).
In the next Section, we briefly describe the corresponding imaging and \Sauron\
datasets and its analysis.

\subsection{Photometric parameters}
\label{sec:photo}
Various parameters were extracted from the imaging data at our disposal.
We first derived radial profiles for all standard variables such as ellipticity
$\epsilon$, position angle PA$_{phot}$, semi-major axis $a$ and disciness/boxiness 
as given by $a_4$ from a best fit ellipse routine, making use of the adapted functionality in the kinemetry
routines of \cite{Krajnovic+06}. The moment ellipticity $\epsilon$ 
profiles were computed within radially growing
isophotes via the diagonalisation of the inertia tensor as in e.g. C+07.
This departs from a simple luminosity-weighted average of the ellipticity
profile, which is more strongly biased towards the central values.
For these profiles, we associate the "aperture" radii $R$: for a given
elliptical aperture or isophote with an area $\cal A$, $R$ is defined as the radius of 
the circle having the same area ${\cal A} =\pi {R}^2$. 
These profiles (curves of growth) are then interpolated 
to obtain parameter values at e.g., one (or half) an effective radius~$R_e$
(provided in Table~5 of Paper~I).
To build diagrams together with quantities derived via the \Sauron\
integral-field data (see next Section), we use ellipse and position angle
profiles provided by the photometry and limit the aperture radius $R$ to a
maximum value $R_S$: it is the minimum between the considered radius (e.g., 1~$R_e$) and the radius
$R_{max}$ for which the corresponding ellipse differs in area not more than 15\% from the actual field coverage 
provided by our spectrographic data, with the ellipse itself lying at least 75\% within that field of view. This guarantees that we have both a good coverage in area (85\%) and that the \Sauron\ spaxels reach sufficiently far out with respect to the borders of the considered aperture. Changing these criteria only affects the measured aperture values for a few systems, and does not modify the global results presented here.
These radial profiles were extracted from the available ground-based data. For
most of the galaxies, we relied on the green g band, close to the wavelength
range covered with the \Sauron\ datacubes. For only a few galaxies, when the g band data was not
available or of poor quality, we instead relied on the red r band or even on the \Sauron\ images 
reconstructed directly from the datacubes: the data used for each individual
galaxy is indicated in Table~\ref{tab:LRparam}.

A number of galaxies in our sample exhibit strong bars \citep[e.g. NGC\,936, NGC\,6548, see][hereafter Paper~II]{Krajnovic+11}. When the
galaxy is viewed at rather low inclination (close to face-on), the bar strongly influences the
measured position angle (as well as the ellipticity), implying a strong
misalignment between the photometric and kinematic major-axes.
The (stellar) kinematic major-axis is an excellent indicator of the line of nodes of
a disc galaxy, even when the galaxy hosts a relatively strong bar, and this
kinematic axis generally coincides with the outer photometric major-axis outside the bar, 
where the light distribution is dominated by a disc. The measured flattening
does however not properly reflect the instrinsic flattening of the galaxy when
measured in the region of the bar. In galaxies with obvious bars, such as NGC\,936, NGC\,3400, NGC\,3412, 
NGC\,3599, NGC\,3757, NGC\,3941, NGC\,4262, NGC\,4267, NGC\,4477, NGC\,4608, NGC\,4624, NGC\,4733, NGC\,4754, NGC\,5473, 
NGC\,5770, NGC\,6548, UGC\,6062, we therefore use the global kinematic
position angle, as derived from the two-dimensional \Sauron\ stellar kinematics, 
with the moment ellipticity value from the outer parts of the galaxy 
(outside the region influenced by the bar; values provided in Paper~II), 
both for the derivation of e.g.,
$\lambda_R$, and for all plots of the present paper.

\subsection{The \Sauron\ data} 

\begin{figure}
\centering
\epsfig{file=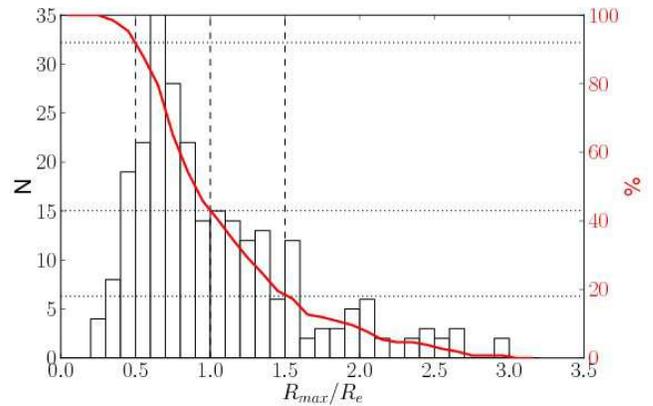, width=\columnwidth}
\caption{Histogram of the maximum aperture radius $R_{max}$ covered by the \Sauron\
observations of all 260 \atlas\ galaxies (normalised by $R_e$). The red line
shows the corresponding cumulative function (right vertical scale) for galaxies with $R > R_{max}$: we
cover about 92\%, 43\% and 18\% at $R_e/2$, $R_e$ and $1.5\,\,R_e$,
respectively, as indicated by the vertical/horizonthal (dashed/dotted) lines.}
\label{fig:Rmax}
\end{figure}
\begin{figure}
\centering
\epsfig{file=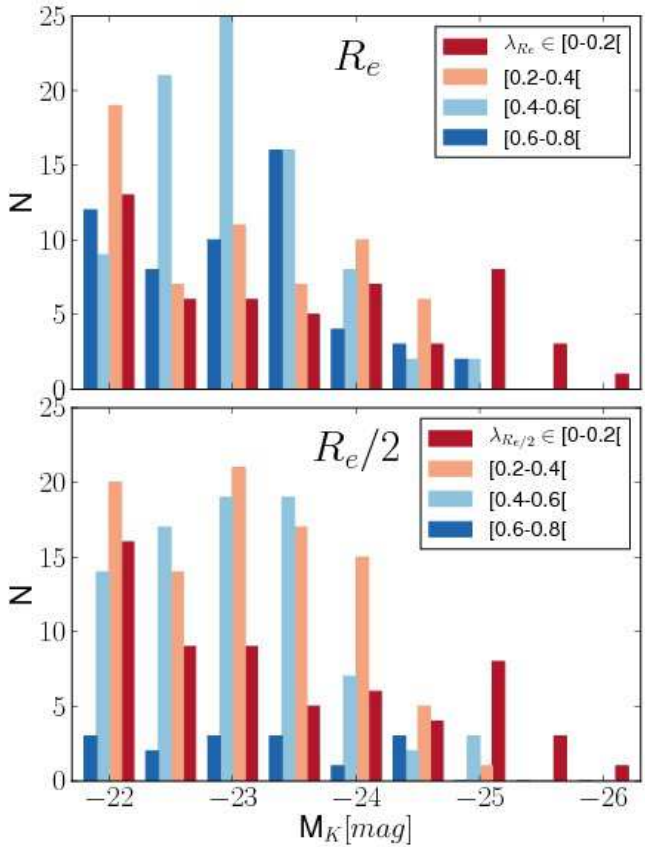, width=\columnwidth}
\caption{Histograms of K band luminosities for all 260 \atlas\ galaxies, in bins
of ${\lambda_R}_e$ (red : [0 -- 0.2]; light red : [0.2 -- 0.4]; light blue : [0.4 -- 0.6]; blue
: [0.6 -- 0.8]). The top panel uses $\lambda_R$ values derived for an aperture
radius of 1~$R_e$, the bottom panel for an aperture radius of $R_e/2$.}
\label{fig:histLum}
\end{figure}
The \Sauron\ integral-field spectrograph (Integral Field Unit, hereafter IFU) 
has been extensively used at the Cassegrain focus of the William Herschel Telescope since 1999 \citep{Bacon+01}. 
All observations were conducted 
using the low spatial resolution mode which provides
a field of view of about 33\arcsec$\times$41\arcsec and a 
spatial sampling of 0\farcs94$\times$0\farcs94. The narrow spectral range allows the user to probe a few stellar absorption and ionised gas emission lines with a spectral resolution of about 4 Angstr\"oms (FWHM). 

All data reduction was performed using the dedicated \XSauron\ software wrapped
in a scripted pipeline. A set of 64 galaxies included in the \atlas\ sample were observed
prior to the mounting of the volume phase holographic (VPH) grating
\citep[mostly from the original \Sauron\ survey, see][for details]{deZeeuw+02}.
For these galaxies, we obtained an average of 2 hours on source sometimes
following a mosaicing strategy to cover the targets with the largest effective
radii. The spectral resolution attained for these galaxies is about 4.2
Angstr\"oms FWHM. For most of the 196 remaining targets, we
integrated 1 hour on source centred on the object, including two (slightly
dithered) 30mns exposures: only when the source was extended did we allow
for a mosaic of 2 fields to attempt to fully cover the region within 1~$R_e$, 
with two 30mns exposure for each field. 
The orientation of the \Sauron\ field was adapted to
each target to optimise the coverage of the galaxy taking into account its
apparent photometric flattening. The spectral resolution attained for these 196
galaxies is about 10\% better (due to the use of the VPH grating) and reaches
3.9 Angstr\"oms FWHM.

The 260 final merged datacubes (with 0\farcs8$\times$0\farcs8 rebinned spaxels) were then analysed using a common analysis pipeline, 
and using a minimum signal-to-noise threshold of 40 for the adaptive binning \citep{CC03}.
Gas and stellar kinematics were extracted via a pPXF algorithm \citep{CappEms04}
with a library of stellar templates as in \cite{Emsellem+04}, but adopting here 
the MILES library \citep{Sanchez+06} and an optimised template per galaxy (see Paper~I for details).

We derived $\lambda_R$ and $V/\sigma$ from growing effective apertures, as
in E+07, following the ellipticity and position angle profiles obtained from the photometry,
or from the constant values (kinemetric axes and moment
ellipticity from the outer part) for galaxies with obvious bars (see
Sect.~\ref{sec:photo}). Using
two-dimensional spectroscopy, the expression for $\lambda_R$ as given by:
\begin{equation}
\lambda_R \equiv \frac{\langle R \, \left| V \right| \rangle }{\langle R \, \sqrt{V^2 + \sigma^2} \rangle}
\, ,
\end{equation} 
transforms into
\begin{equation}
\label{eq:sumLambda}
\lambda_R = \frac{\sum_{i=1}^{N_p} F_i R_i \left| V_i \right|}{\sum_{i=1}^{N_p} F_i R_i \sqrt{V_i^2+\sigma_i^2}} \, ,
\end{equation}
where $F_i$, $R_i$, $V_i$ and $\sigma_i$ are the flux, circular radius, velocity and
velocity dispersion of the i$^{th}$ spatial bin, the sum running on the $N_p$
bins.  Considering the 
signal-to-noise threshold used here, we expect a typical positive bias for values of
$\lambda_R$ near zero (see Appendix A of E+07) in the range [0.025- 0.05].

In the following, we will use ${\lambda_R}_e$ and $(V/\sigma)_e$ to denote
values measured at 1~$R_e$, and ${\lambda_R}_{e/2}$ and $(V/\sigma)_{e/2}$ for values at $R_e / 2$.
In Fig.~\ref{fig:Rmax}, we provide an histogram of the maximum available effective 
aperture sizes from our \Sauron\ dataset for the \atlas\ sample: we cover an
aperture of at 1~$R_e$ or larger for 43\% of our sample and $R_e/2$ for 92\% of all
\atlas\ galaxies. Note that 18\% are covered up to at least $1.5 \,R_e$.

Finally, accurate dynamical masses $M_{\rm dyn}$ were derived via
Multi-Gaussian Expansion \citep{Emsellem+94} of the galaxies' photometry
followed by detailed Jeans anisotropic dynamical models
\citep{Cappellari08,Scott+09} of the \atlas\ \Sauron\ stellar kinematics
\citep{Cap+10}. This
mass represents $M_{\rm dyn} \approx 2 \times M_{1/2}$, where $M_{1/2}$ is the
total dynamical mass within a sphere containing half of the galaxy light.


\section{A first look at the apparent angular momentum of \atlas\ galaxies}
\label{sec:LambdaR}

\subsection{Velocity structures and $\lambda_R$}
\label{sec:vel}

With the 260 stellar velocity maps from the \atlas\ sample, we probe the whole
range of velocity structures already uncovered by E+07: regular disc-like velocity fields
(e.g., NGC\,4452, NGC\,3530), kinematically distinct cores (e.g., NGC\,5481,
NGC\,5631) or counter-rotating systems (NGC\,661, NGC\,3796), twisted
velocity contours (e.g., NGC\,3457, NGC\,4552), 
sometimes due to the presence of a bar (e.g., IC\,676, NGC\,936).
We also observe a few more galaxies with two large-scale counter-rotating disc-like
components, as in NGC\,4550 \citep{Rubin+92}, e.g., IC\,0719 or NGC\,448. IC\,719 exhibits in fact two
velocity sign changes along its major-axis, and NGC\,4528 three sign changes.
Only a few galaxies have noisy maps or suffer from systematics, e.g., NGC\,1222, UGC\,3960, 
or PGC\,170172, due to the low signal to noise ratio of the associated
datacubes or from intervening structures (e.g., stars). 
The reader is refered to Paper~II for further details on the
kinematic structures present in galaxies of the \atlas\ sample.

The \atlas\ sample of ETGs cover ${\lambda_R}_e$ values from 0.021, 
with M\,87 (NGC\,4486), therefore consistent with zero apparent angular
momentum within the \Sauron\ field of view, and 0.76 for NGC\,5475 a flattened
disc-like galaxy. Other galaxies with low ${\lambda_R}_e$
values and stellar velocity fields with nearly zero velocity amplitude (within
the noise level) are NGC\,3073, NGC\,4374, NGC\,4636, NGC\,4733, NGC\,5846, and NGC\,6703. 
Of these six, NGC\,4374, NGC\,4636, and NGC\,5846 are nearly round, 
massive galaxies with a mass well above $10^{11}$~\Msun, and strong X-ray emitters. 
NGC\,4636 shows a very low amplitude velocity field and a
barely detectable kinematically distinct component. There are two more
galaxies with significantly non-zero values of ${\lambda_R}_e$ ($\sim 0.1$) 
but no detectable rotation, the
relatively high ${\lambda_R}_e$ values being due to larger uncertainties in the
kinematics: NGC\,3073 and NGC\,4733. Along with NGC\,6703, these stand out as galaxies
with no apparent rotation, a mass below $10^{11}$~\Msun\, and an effective
radius smaller than 3~kpc: these are very probably nearly face-on disc-galaxies
(NGC\,4733 being a face-on barred galaxy).

If we use the previously defined threshold separating Slow and Fast Rotators,
namely ${\lambda_R}_e = 0.1$, we count a total of 23 galaxies below that limit in our sample,
a mere 9\% of the full \atlas\ sample. This is to be compared with one fourth 
(25\%) of galaxies below that threshold found in E+07 
within the representative \Sauron\ sample of 48 early-type galaxies
\citep{deZeeuw+02}. With \atlas\ we
are covering a complete volume-limited sample, more than five times larger than the original
\Sauron\ sample, but we less than double the number of such slowly rotating
objects. This is expected considering that such galaxies tend to be on the high luminosity
end (E+07, C+07). A volume-limited sample includes far more galaxies in the low-luminosity range
than the \Sauron\ representative sample which had a flat luminosity
distribution. Our \atlas\ observations confirm the fact that slow and
fast rotators are not distributed evenly in absolute magnitude: fast-rotating systems brighter 
than M$_K$ of -24 are rare, while a third of all galaxies having ${\lambda_R}_e \leq 0.1$ are in this range.
This is emphasised in Fig.~\ref{fig:histLum} where
the K band luminosity histograms for galaxies in bins of ${\lambda_R}_e$ are presented. 
This trend could be due to an incomplete field coverage of the brighter galaxies with large $R_e$ 
when $\lambda_R$ increases outwards (we reach 1~$R_e$ for only 
about 42\%, see Fig.~\ref{fig:Rmax}, and this is obviously biased toward the fainter end of galaxies), 
However, this trend is still present when using $\lambda_R$ at $R_e / 2$ 
(bottom panel of Fig.~\ref{fig:histLum}) 
for which we have about 92\% of all galaxies properly covered. 

Among the 23 galaxies with ${\lambda_R}_e \leq 0.1$ in the \atlas\ sample, 6 have no detected rotation
(3 of them being very probably face-on disc-like systems, see above), 2 have twisted or
prolate-like isovelocity contours (NGC\,4552 and NGC\,4261), and out of the remaining 15 others, 14 have
Kinematically Distinct Cores (KDCs, see Paper~II), confirming the 
claim made in E+07 that most ETGs with low $\lambda_R$ values have KDCs. 
\begin{figure*}
\centering
\epsfig{file=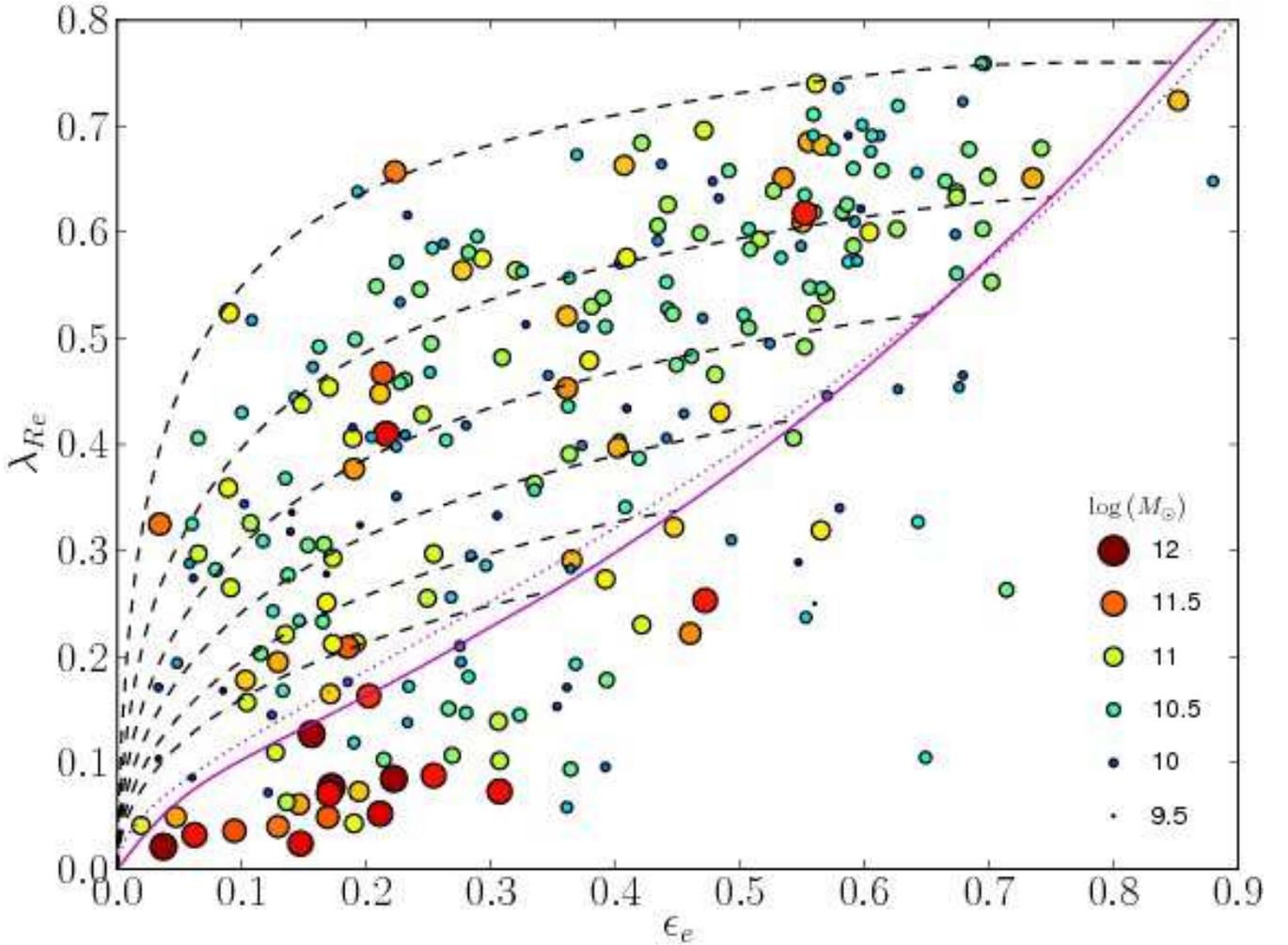, width=\hsize}
\caption{${\lambda_R}_e$ versus ellipticity $\epsilon_e$ for all 260 \atlas\ galaxies.
The colour and size of the symbols are associated with the mass of each galaxy,
as indicated at the bottom right of the panel. The dotted magenta line
show the edge-on view for ellipsoidal galaxies integrated up to infinity 
with $\beta = 0.70 \times \epsilon$, as in C+07. The solid magenta line
is the corresponding curve restricted to an aperture at 1~$R_e$ and for
$\beta = 0.65 \times \epsilon$ (see text for details). The black dashed lines correspond to the location
of galaxies with intrinsic ellipticities $\epsilon_{intr} = 0.85, 0.75, 0.65, 0.55, 0.45, 0.35$ 
along the relation given for an aperture of 1~$R_e$ with 
the viewing angle going from edge-on (on the relation) to face-on (towards the
origin).}
\label{fig:LREpsMass}
\end{figure*}
\begin{figure}
\centering
\epsfig{file=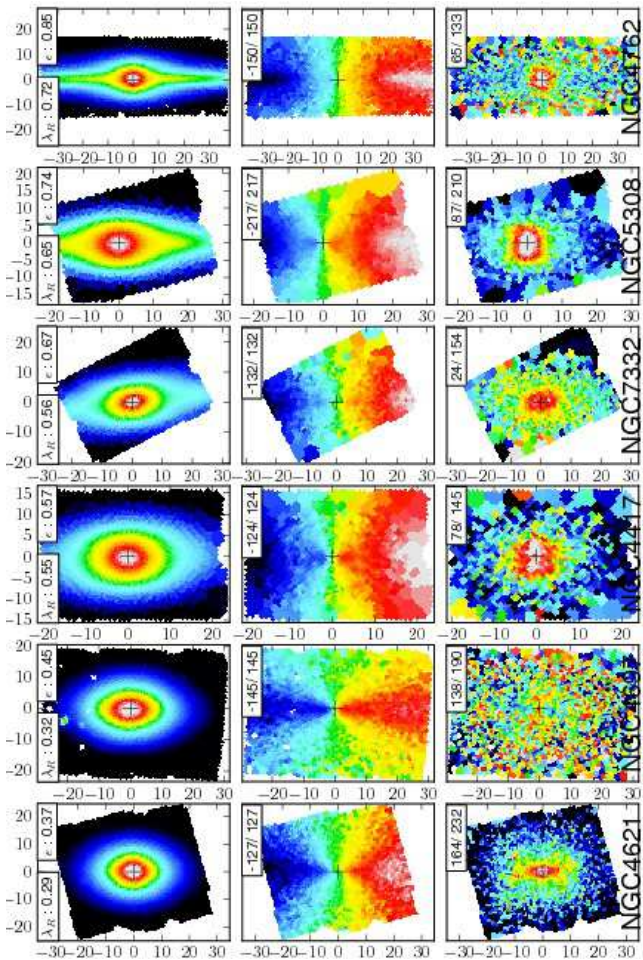, width=\columnwidth}
\caption{Reconstructed images and first two stellar velocity moment maps (velocity
and velocity dispersion) for 6 galaxies selected near the 
ellipticity-anisotropy relation (magenta line, see text).
The apparent ellipticity is decreasing from top to bottom.}
\label{fig:lineVmap}
\end{figure}
\subsection{\atlas\ galaxies in a $\lambda_R$-$\epsilon$ diagram}

The combination of the measured ${\lambda_R}_e$ values with the apparent
flattening $\epsilon_e$ holds important clues pertaining to the intrinsic
morphology and dynamics of ETGs, as shown in E+07. 
In Fig.~\ref{fig:LREpsMass}, we provide a first glimpse at the distribution of
galaxies in such a $\lambda_R$-$\epsilon$ diagram for an aperture radius of 1~$R_e$.

A more standard approach includes the use of $(V/\sigma)$ as a probe for the stellar 
kinematics of galaxies. In C+07, it has been shown that there seems to be a broad trend between the anisotropy
of ETGs, parameterized\footnote{$\beta$ is the anisotropy parameter simply defined as $1 - \sigma_z^2 /
\sigma_R^2$ for a steady-state system where $\sigma_{R,z}$ are the cylindrical components 
of the stellar velocity dispersion.} with $\beta$, and their intrinsic (edge-on) ellipticity
$\epsilon_{intr}$. Fast rotators were found to be generally distributed on 
the $(V/\sigma)$-$\epsilon$ diagram within the envelope traced by the edge-on
relation $\beta = 0.7 \times \epsilon_{intr}$ (from the analytic formula of \citet{Binney05})
and by its variation with inclination (Fig.~11 of C+07).
This analytic relation is nearly identical to the one $\beta = 0.65 \time
\epsilon$, which includes aperture integration within 1~$R_e$ (Appendix~B).
Since $V/\sigma$ and $\lambda_R$ of simple ellipsoidal systems (with constant
anisotropy) can be linked via a relatively simple formula (see Appendix~B), we
can translate these $\beta$-$\epsilon_{intr}$ relations for $\lambda_R$ and provide
the corresponding curves in a $\lambda_R$-$\epsilon$ diagram.
These relations are shown in Fig.~\ref{fig:LREpsMass} for edge-on galaxies 
(dashed and solid magenta lines) as well as the effect of inclination (dashed black lines, only for 
the relation integrated within 1~$R_e$). 

We first confirm that most of the galaxies with ${\lambda_R}_e$ values significantly above 0.1
are located above (or at the left) of the magenta line in Fig.~\ref{fig:LREpsMass}.
The dashed line at $\epsilon_{intr} = 0.85$ also provides a convenient upper envelope of the
galaxies in our sample. This beautifully confirms the predictions made in C+07, using
only a small set of targets, and reveals important characteristics of the
internal state of early-type galaxies, which will be further discussed in
Sect.~\ref{sec:MC}.

The majority of galaxies above the magenta line are consistent with intrinsic ellipticities 
between 0.55 and 0.85, with only very few galaxies below $\epsilon_{intr} = 0.35$.
The stellar velocity maps of these fast rotating objects, illustrated
in Fig.~\ref{fig:lineVmap} with 6 examples of galaxies 
from low ($\sim 0.35$, bottom), to high ($\sim 0.85$, top) ellipticities along the magenta line,
are regular and show disc-like signatures (e.g., pinched isovelocity contours), strongly contrasting with the
complex kinematic features observed for galaxies with low ${\lambda_R}_e$ values, as
discussed in Sect.~\ref{sec:vel}: this is objectively quantified in Paper~II.
We here probe from very flattened edge-on cases dominated by a thin disc
component at the top, to relatively fatter objects like NGC\,4621
at the bottom. All 6 galaxies exhibit a clear sign of a disc-like component either from 
flattened isophotes and/or pinched iso-velocities, confirming the fact that they
cannot be viewed far from edge-on.

The distribution of galaxies in the \atlas\ sample in the $\lambda_R$-$\epsilon$
plane also reveals a rather well-defined upper envelope: as ellipticity decreases,
the maximum apparent angular momentum decreases accordingly. It can 
be understood by looking at Fig.~\ref{fig:LREpsMass} again where the upper black dashed line,
corresponds to an extremely flattened spheroid with $\epsilon_{intr} = 0.85$ (a disc) and an anisotropy
parameter of $\beta = 0.55$ (following the $\beta$-$\epsilon_{intr}$ relation
mentioned above): most galaxies in the \atlas\ sample have lower $\lambda_R$.

This first view at the distribution of the \atlas\ galaxies in
a $\lambda_R$-$\epsilon$ diagram provides a very significant upgrade on
already published samples. We can therefore now proceed by combining the
detailed study of the kinematic structures observed in these galaxies conducted
in Paper~II with such information to deliver a refined criterion for
disentangling Fast and Slow Rotators.

\section{Galaxy classification via stellar kinematics}
\label{sec:slowfast}

\subsection{The importance of shapes}
\label{sec:twofamilies}
We have seen in the previous Section that galaxies with the lowest $\lambda_R$ values 
tend to be more luminous or massive (see Figs.~\ref{fig:histLum} and \ref{fig:LREpsMass}), and
exhibit complex kinematic structures, as opposed to fast-rotating galaxies
with regular velocity fields and disc-like signatures (when viewed near edge-on). 
This confirms the view delineated in E+07 and C+07,
where early-type systems were separated into two families, the
so-called Fast and Slow Rotators. However, it is not clear whether 
a {\em constant} value of $\lambda_R$ (e.g., $\lambda_R = 0.1$ 
as defined in E+07 from a representative sample of 48 galaxies) corresponds
to the best proxy for distinguishing between slow and fast rotators. 

Using the complete \atlas\ sample of 260 galaxies, we can in fact proceed with an improved
criterion, taking into account the fact that two galaxies with the
same apparent angular momentum but very different (intrinsic) flattening must have, by
definition, a different orbital structure. A galaxy with a relatively low value of $\lambda_R$,
e.g., of 0.2, may be consistent with a simple spheroidal axisymmetric system viewed at a
high inclination (near face-on), but this is true only if its ellipticity is 
correspondingly low: a large ellipticity, e.g., $\epsilon=0.4$, would 
imply a more extreme object (in terms of orbital structure or anisotropy), as
shown with the spheroidal models provided in Appendix~B.

\begin{figure}
\centering
\epsfig{file=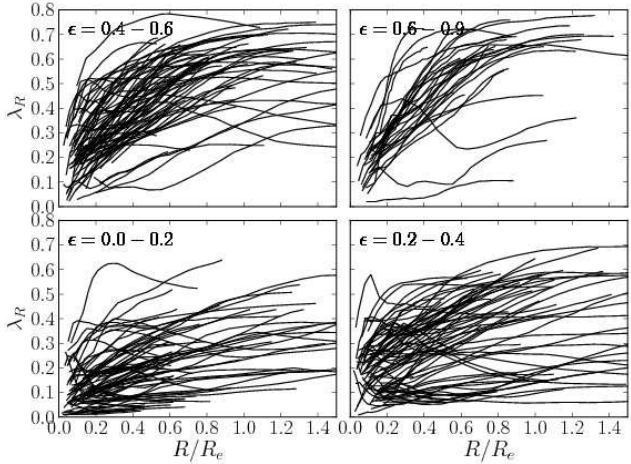, width=\columnwidth}
\caption{$\lambda_R$ profiles for the complete \atlas\ sample of early-type
alaxies, in four bins of ellipticities  $\epsilon_e$ (as indicated in each panel), as a
function of the aperture radius (relative to the effective radius $R_e$).}
\label{fig:LRprof}
\end{figure}

This can be further illustrated by looking at Fig.~\ref{fig:LRprof} where
the radial $\lambda_R$ profiles are shown for all galaxies of the sample
in bins of ellipticities. For low ellipticities, $0 < \epsilon < 0.2$, there is
a rather continuous sequence of profiles with various $\lambda_R$ amplitudes from 0.1 to 0.5 at
1~$R_e$: a rather face-on flattened system would have a profile similar to an inclined mildly-triaxial galaxy.
In the next bin of ellipticities, $0.2 < \epsilon < 0.4$, we start discerning two main groups
of galaxies: the ones with rapidly increasing $\lambda_R$ profiles, most of
these galaxies showing regular and symmetric velocity fields (as in Fig.~\ref{fig:lineVmap}), and those
who have flatter (or even decreasing) profiles up to $\sim R_e/2$ and then start increasing outwards,
again often exhibiting complex velocity maps and distinct central stellar velocity structures.
There are in addition a few galaxies with $\lambda_R$ profiles going up to
$\sim 0.3$ and {\em decreasing} between $R_e / 2.$ and $R_e$.
As we reach the last ellipticity bin (with the highest values), most galaxies have
strongly rising $\lambda_R$ profiles reaching typical values of $\lambda_R \sim 0.5$ 
at $R_e / 2$. The 3 galaxies which have ${\lambda_R}_e$ below 0.4 are 
IC\,719, NGC\,448 and the famous NGC\,4550, all being extreme
examples of disc galaxies with two counter-rotating systems.

A refined scheme to separate slowly and fast rotating galaxies should therefore
take into account the associated apparent ellipticity: we should expect
a higher value of the specific stellar angular momentum for galaxies which are
more flattened or closer to edge-on if these are all intrinsically fast
rotators. Very flattened galaxies with ${\lambda_R}_e$ as low as 0.2 or 0.3 must
already have a rather extreme orbital distribution (or strong anisotropy, see Appendix~B).
In the next Section, we use the completeness of our sample of early-type
galaxies to revisit the question of how to distinguish members of the two main
families of early-type galaxies, namely Slow and Fast Rotators.

\subsection{Kinemetric structures and the link with $\lambda_R$}

The classification of ETGs in Slow and Fast rotators was motivated by the
(qualitative) realisation that galaxies within the \Sauron\ sample of 48 ETGs
\citep{deZeeuw+02} exhibit either regular stellar rotation, showing up as
classical "spider-diagrams", with no significant misalignment between the kinematic 
and photometric axes (excluding the few systems consistent with having no apparent rotation at
all), or complex/irregular stellar velocity maps with twists
and strong misalignment with respect to the photometry \citep{Emsellem+04, PaperXII}. Using $\lambda_R$ as a
proxy to disentangle the two families of objects, it was confirmed that these
two families had distinct structural and dynamical properties (E+07, C+07).

We can now review these results in the context of our complete \atlas\ sample:
this requires an objective assessment of the observed kinematic structures.
The regularity or richness of a velocity map can be defined and more importantly
quantified using kinemetry \citep{Krajnovic+06}. Such an evaluation has thus
been conducted in Paper~II systematically for all 260 \atlas\ galaxies.
The (normalised) amplitude of the 5$^{th}$ harmonic kinemetric term $k_5$ ($k_5/k_1$)
can for example be used to find out whether or not a velocity field has
iso-velocity contours consistent (in azimuth) with the cosine law expected from pure disc
rotation. The fact that the velocity map of a galaxy follows the cosine law does not
directly imply that it is a pure-disc system, only that its line-of-sight velocity
field looks similar to one of a two-dimensional disc. 
Using a threshold of 4\% for $k_5/k_1$, galaxies with 
or without regular velocity patterns have been labelled in Paper~II 
as Regular and Non-Regular Rotators, respectively, and provided a
detailed and quantified account of observed velocity structures.
Galaxies such as NGC\,3379 or NGC\,524 exhibit low amplitude rotation but are consistent with being regular rotators, 
while galaxies like NGC\,4406 or NGC\,4552 are clearly non-regular rotators (see Fig.~B1, B5 and B6 of
Paper~II) even though these are rather round in projection.  

All 260 galaxies of the \atlas\ sample are shown with symbols for regular and
non-regular rotators in Fig.~\ref{fig:LRVS_NDR}: we plot both measured ${V/\sigma}_e$
and ${\lambda_R}_e$ as functions of ${\epsilon}_e$. The expected locations
for systems with $\beta = 0.7 \times \epsilon$, where $\beta$ is the anisotropy
parameter are presented as dotted lines, the magenta line representing edge-on
systems and the dashed black line showing the effect of inclination for an
intrinsic (edge-on) ellipticity of 0.82. These curves are calculated following the
formalism in \cite{Binney05}, and the specific values are identical to the ones 
defined in C+07. We also provide similar relations for $\beta = 0.65 \times \epsilon$
as in Fig.~\ref{fig:LREpsMass}, but this time taking into account the effect of a limited
aperture (1~$R_e$). 

We observe in Fig.~\ref{fig:LRVS_NDR} that regular rotators tend to have $V/\sigma$
or $\lambda_R$ values above the magenta line. This clearly confirms the trend
emphasised in C+07 already obtained with a significantly smaller sample. The \atlas\
sample clearly extends this result to galaxies at higher
ellipticities and $V/\sigma$ or $\lambda_R$ values. The second obvious and
complementary fact is that non-regular rotators cluster in the lower part of the diagrams, and below the magenta line. 
Overall, this shows that objects with or without specific kinematic features 
in their velocity or dispersion maps tend to be distributed on either
side of the relation illustrated by the magenta line.

This strongly suggests that the regularity of the stellar velocity pattern are closely related to the Slow
and Fast rotators classes, as defined in E+07, and that
the \atlas\ sample of 260 galaxies provides the first view of a complete sample
of ETGs, expanding on the perspective derived from the original \Sauron\ sample
of 48 galaxies. We do not expect a one to one relation between non-regular
rotators and Slow rotators on one hand, and regular rotators and fast rotators on the other hand, because e.g., any
departure from a regular disc-like rotation automatically qualifies a
galaxy as a non-regular rotation pattern. However, apart from atypical cases such as unrelaxed merger
remnants or galaxies with strong dust features, we may expect that such a 
link holds. We now examine how to best separate these two families of ETGs.

\subsection{Slow and Fast rotators}
\begin{figure}
\centering
\epsfig{file=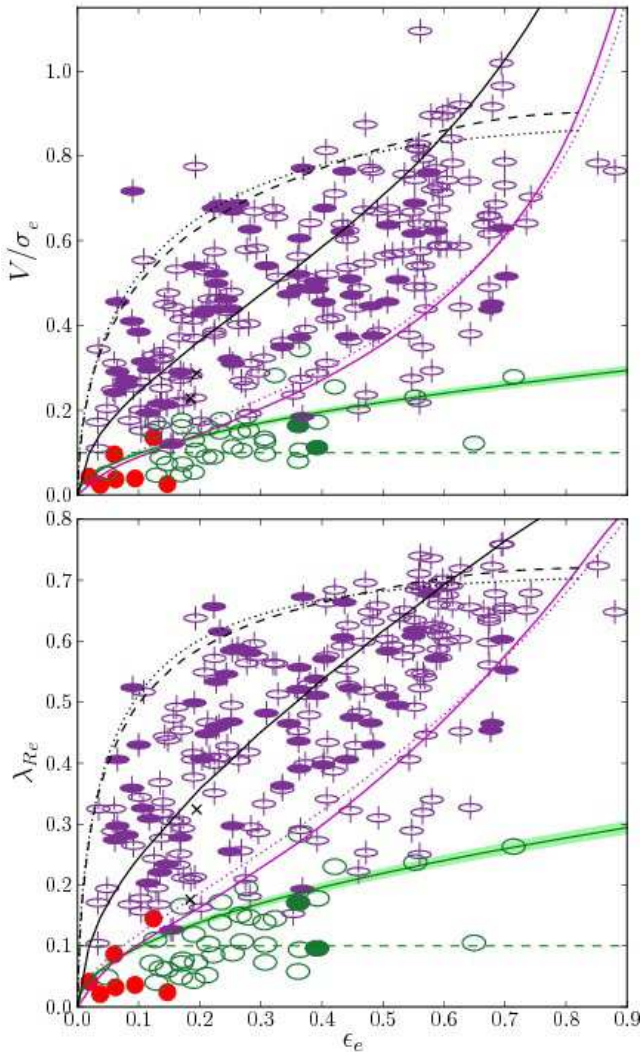, width=\columnwidth}
\caption{Measured ${V/\sigma}_e$ (top) and ${\lambda_R}_e$ (bottom) values versus the
ellipticity ${\epsilon}_e$ within an aperture of 1~$R_e$.
The magenta lines are as in Fig.~\ref{fig:LREpsMass}, and the black dotted and
dasked lines represent the location of galaxies with an intrinsic ellipticity $\epsilon=0.82$
when going from an edge-on to a face-on view. The solid black line corresponds
to isotropic oblate systems viewed edge-on. The solid green line is $0.31
\times \sqrt{\epsilon}$. 
Red circles are galaxies with no apparent rotation, green ellipses and purple
symbols are for non-regular rotators and regular rotators, respectively. Filled symbols correspond to
galaxies with bars. Regular and non-regular rotators are better separated using \lr\ than
$V/\sigma$.}
\label{fig:LRVS_NDR}
\end{figure}

The $V/\sigma$ diagram (Fig.~\ref{fig:LRVS_NDR}) shows, independently from any refined
criterion to disentangle regular and non-regular rotators,
significantly higher overlap between the two populations both at low and high ellipticities. 
The difference between $V/\sigma$ and $\lambda_R$ may not, however, be that obvious just
looking at Fig.~\ref{fig:LRVS_NDR}. For simple oblate models, as illustrated in Appendix~B,
there is a tight correlation between these two quantities (see Fig.~\ref{fig:LRVS}).
This is not the case anymore when the galaxy exhibits more complex kinematics, with
e.g. a rapidly rotating inner part and a slowly-rotating outer part.
Two galaxies with apparent dynamics as different as NGC~5813 and NGC~3379 
have similar $V/\sigma$ values, although the latter is a regular rotator while the former 
is an non-regular rotator with a clear stellar KDC (see E+07). In this context, $\lambda_R$ is a 
better discriminant and this occurs because the weighting of stellar rotation 
depends both on the observed flux and on the size (radius) of the structure. 
While the difference may not be dramatic and would not impact the majority of
ETGs (since most have regular stellar velocity fields), it becomes relevant when
considering classes of galaxies for which we observe differences in the observed
kinematic features. This has motivated the use of ${\lambda_R}$ 
which also directly relates to the apparent angular momentum of the stars (E+07).

Before we refine the above-mentioned criterion, we need to emphasise again the clear 
trend observed in the $\lambda_R$-$\epsilon$ diagram (bottom
panel of Fig.~\ref{fig:LRVS_NDR}): galaxies have on average increasing values of 
$\lambda_R$ as the ellipticity increases, and this is valid also for non-regular
rotators alone. As mentioned in Sect~\ref{sec:twofamilies},
at constant $\lambda_R$ value, the anisotropy increases with higher
ellipticities, and we therefore need to define a threshold for slow/fast rotators
which depends on and increases with ellipticity.

We considered several possibilities, including scaled-down versions of the predicted ${\lambda_R}-\epsilon$ 
relation for isotropic axisymmetric systems or of the magenta lines.
The first one (isotropy being used as a reference) does a reasonable job 
at separating regular and non-regular rotators,
with a scaling factor of $\sim 0.4$ and would naturally connect our study with already published work.
The fact that galaxies with low ${V/\sigma}$ or $\lambda_R$ exhibit different observed
properties is certainly not a new result, and was illustrated and emphasised 
in a number of key studies \citep[e.g.,][]{Bender88, KB96} using the ratio
${V/\sigma}_{\star}$ between the measured ${V/\sigma}$ 
and the predicted value from an oblate isotropic rotator ${V/\sigma}_{iso}$
\citep[see e.g.][]{Davies+83}. ${V/\sigma}_{\star}$ has thus sometimes
been used as an indication of an anisotropic dispersion tensor: 
this view is in fact misleading as a constant value for ${V/\sigma}_{\star}$ does imply
an increasing anisotropy with increasing flattening (C+07). This statement is also valid 
for ${\lambda_R}_{\star} = {\lambda_R} / {\lambda_R}_{iso}$ obviously for the same reason.

More importantly, C+07 have shown that galaxies can generally not
be considered as isotropic (see Fig.~\ref{fig:LRVS_NDR}). Global anisotropy increases with the
intrinsic flattening, therefore using isotropic rotators as a reference for
flattened galaxies would not be appropriate. Galaxies in the \atlas\ sample appear
to be distributed around the isotropy line in Fig.~\ref{fig:LRVS_NDR}, but
this relation is derived for {\em edge-on} systems, and galaxies 
should span the full range of inclinations with roughly as
many galaxies above and below an inclination of 60\degr. 
Scaling of the magenta line would in this context
be more appropriate, although it clearly has a similar drawback: it is defined
for intrinsically edge-on systems, does not follow the variation of $\lambda_R$
and $\epsilon$ due to inclination effects, and therefore does not perform well in disentangling regular rotators
from non-regular ones.

After considering various possibilities, we finally converged on what we
believe is the simplest proxy which can properly account for the two observed families, minimising the
contamination on both sides. We therefore fixed the threshold for $\lambda_R$
to be proportional to $\sqrt{\epsilon}$ with a scaling parameter $k_{FS}$ which depends
on the considered apertures, namely:
\begin{equation}
   \label{eq:FS}
   {\lambda_R}_e = \left( 0.31 \pm 0.01 \right) \times \sqrt{\epsilon_e} 
\end{equation}
\begin{equation}
   \label{eq:FS2}
   {\lambda_R}_{e/2} = \left( 0.265 \pm 0.01\right)   \times \sqrt{\epsilon_{e/2}}
\end{equation}
The different values of $k_{FS}$ for these two apertures obviously follow the observed 
mean ratio between ${\lambda_R}_e$ and ${\lambda_R}_{e/2}$ (see
Appendix~B). Here, $\lambda_R$ and $\epsilon$ are {\em measured} values:
formal errors for these are very small (because these
parameters are computed using many independent spaxels). 
Uncertainties in the measurements of $\lambda_R$ are thus mostly affected by systematic errors
in the stellar kinematic values and are difficult to assess. 
The quoted ranges in Eqs.~\ref{eq:FS} and \ref{eq:FS2} ($[0.30-0.32]$ and $[0.255-0.275]$ for 
$R_e$ and $R_e/2$, respectively) are therefore only indicative of the difficulty
in defining such empirical thresholds.
The relation for an 1~$R_e$ aperture is shown as a solid green line (the filled area showing the
quoted ranges) in Fig.~\ref{fig:LRVS_NDR}: 
it performs well in its role to separate galaxies with regular and non-regular velocity patterns,
and does slightly better with $\lambda_R$ than with $V/\sigma$ which shows a 50\% increase of misclassified objects 
(and a larger number of non-regular rotators above the magenta line).
The two non-rotators which are above the green line (NGC\,3073 and NGC\,4733; red circles)
are in fact very probably face-on fast-rotating galaxies. 

Equations~\ref{eq:FS} and \ref{eq:FS2} can therefore be used to define Fast and Slow
rotators, but as for any classification scheme, we need to define
a scale at which to apply the criteria: this is further examined in the next Section.

\subsection{The importance of a scale}
\label{sec:scale}

Large-scale structures ($\sim 2 R_e$ and beyond) are certainly important to understand the formation and
assembly history of galaxies, and reveal e.g., signatures such as faint tidal structures or streams. 
However, we are interested here in probing the central regions of early-type galaxies,
at the depth of their potential wells where metals are expected to have accumulated or been produced, 
and where stellar structures should have had time to dynamically relax: using
apertures of one effective radius follows these guidelines and, moreover, covers
about 50\% of the stellar mass.

Stellar angular momentum is generally observed to increase at large radii (E+07), 
even for slow rotators, and this is also valid for most \atlas\ galaxies (Fig.~\ref{fig:LRprof}).
The majority of fast rotating galaxies reach close to their maximum $\lambda_R$ values between
$R_e/2$ and $R_e$: beyond that radius, the profiles often smoothly bend and reach a
plateau. Many galaxies with low central apparent angular momentum also show 
${\lambda_R}_{e}$ increasing values outside $R_e/2$. 
It is also true that most kinematically distinct components observed 
in our \atlas\ sample have radii smaller than $R_e/2$. And for all the 
detected KDCs in our sample, the maximum radius covered is at least 50\% larger 
than the radius of the KDC itself. 
\begin{figure}
\centering
\epsfig{file=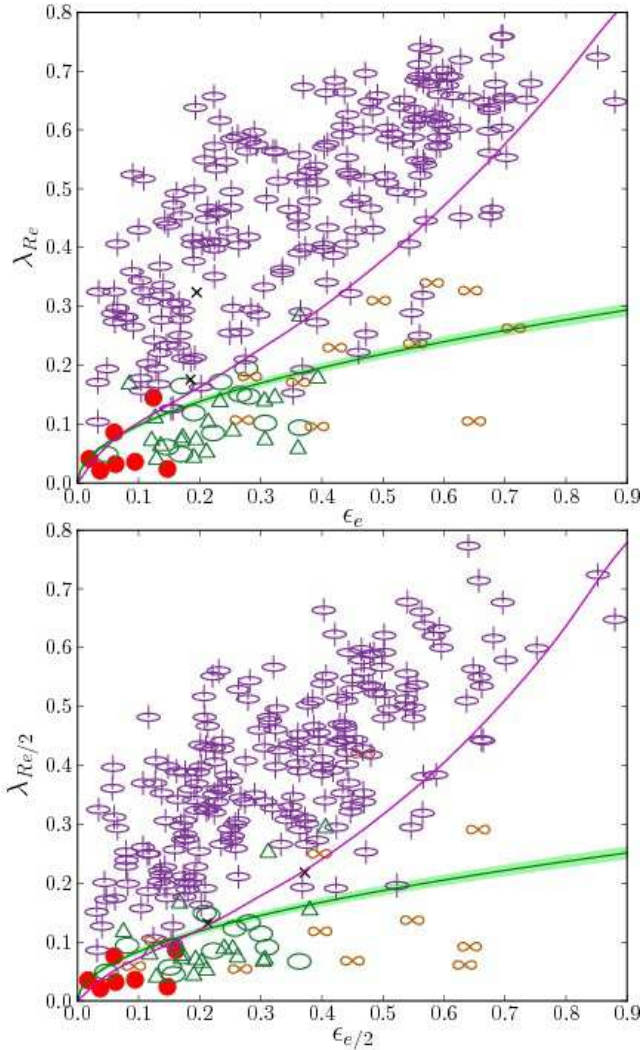, width=\columnwidth}
\caption{As in the bottom panel of Fig.~\ref{fig:LRVS_NDR}, but here for $\lambda_R$ within apertures
of $R_e$ (top panel) and $R_e /2$ (bottom) panel. Symbols correspond to
different kinematic groups (see text for details): red circles are non-rotators,
green ellipses are for non-regular rotators without any specific kinematic feature, green triangles
are galaxies with KDCs, orange lemniscates are 2$\sigma$ galaxies (galaxies
with two counter-rotating flattened stellar components), purple
symbols are regular rotators and black crosses are 2 galaxies which could not be
classified.}
\label{fig:LRLR2_Flag}
\end{figure}
This suggests that both apertures of $R_e / 2$ and $R_e$ could serve as
reference scales to define Slow and Fast Rotators (SRs and FRs, respectively). Using apertures smaller than
$R_e/2$ is not advised, as this would make the measurements more dependent on
instrumental characteristics and observation conditions, and would probe 
only the core regions. To compare how our criterion performs for 1~$R_e$
and $R_e / 2$, we will use a more detailed description of the kinematic features 
present in the maps: this will help interpreting the observed similarities and differences.

In Paper~II, the regular/non-regular rotator types, which globally define
the observed velocity structure of a galaxy (in combination
with key features observed in both the velocity {\em and} velocity dispersion
maps and analysed via kinemetry), have been used to define five kinematic groups:
\begin{itemize}
   \item {\em Group a}: galaxies with no apparent rotation, or non-rotators (7 members);
   \item {\em Group b}: galaxies with non regular velocity pattern (non-regular
      rotators) but without any specific kinematic feature (12 members);
   \item {\em Group c}: galaxies with Kinematically Distinct or Counter-rotating Cores (19 members);
   \item {\em Group d}: galaxies with two symmetrical off-centred stellar velocity dispersion peaks (11 members);
   \item {\em Group e}: galaxies with regular apparent rotation (regular rotators) and with or
      without small minor-axis kinematic twists (209 members).
\end{itemize}
Galaxies of groups {\em a} to {\em d} are mostly non-regular rotators, while most of the galaxies
in group {\em e} are regular rotators. Galaxies of {\em group d}, also called `2$\sigma$'
galaxies due to the appearance of their velocity dispersion maps, 
are interpreted as systems with two counter-rotating 
flattened stellar components which can have various relative luminosity
contributions. This includes galaxies such as the well-known NGC\,4550
\citep{Rubin+92}, which is made of two counter-rotating discs of nearly equal
light \citep[][C+07]{Rix+92}, and for this reason ends up as a slow
rotator, or other cases like NGC\,4473, which has a smaller amount of
counter-rotating stars (C+07) and thus appears as a fast rotator,
or newly discovered objects like NGC\,4528 (Paper~II). 
Among our sample of 260 objects, two galaxies could not be classified due
to the low signal-to-noise ratio of the extracted kinematics.

In Fig.~\ref{fig:LRLR2_Flag}, we now re-examine the $\lambda_R$ values for our \atlas\ sample in the light
of these 5 {\em kinematic groups} for both apertures of $R_e$ and $R_e / 2$. As
expected, both the ellipticity $\epsilon$ and $\lambda_R$ values are smaller
within $R_e/2$: $\lambda_R$ is generally an increasing function of radius
and going inwards we tend to shift away from a large-scale disc structure when
present (the median of our observed  ${\lambda_R}_e / {\lambda_R}_{e/2}$ values
is $\sim 1.17$; see also Appendix~B). All results previously mentioned within an
aperture of 1~$R_e$ are confirmed with a smaller one ($R_e/2$). The
galaxies of {\em kinematic groups  a, b, c, d} seem to nicely cluster below the
green lines defined in Eqs.~\ref{eq:FS} and \ref{eq:FS2}, and the resulting separation of these
targets from the kinematic {\em group e} (regular rotator) is marginally sharper within
$R_e /2$: there are no galaxies of {\em group e} below the green line and 
most 2$\sigma$ galaxies are now below the defined threshold for slow rotators. 

Eqs.~\ref{eq:FS} and \ref{eq:FS2} thus provide excellent (and simple) proxies to discriminate 
between galaxies of {\em kinematic groups a, b, c, d} 
and {\em group e}, or conversely between regular and non-regular rotators,
with only one object with regular disc-like stellar velocity maps below the line 
(NGC\,4476) at 1~$R_e$ and none for the smaller aperture. In fact, NGC\,4476
seems to be a bona fide {\em group d} galaxy, but with an usually large inner
counter-rotating component (Alison Crocker, priv. communication).

Using these criteria, there are 36 Slow Rotators (SRs) out of 260 for an 
aperture of $R_e$ and 37 for $R_e/2$,
with 30 in common for both apertures. All targets which change class going
from $R_e/2$ to $R_e$ are well covered with the available \Sauron\ field of
view, and this is therefore not an effect of spatial coverage. Galaxies which are FRs at
$R_e/2$ and SRs at $R_e$ are NGC\,4476, NGC\,4528, NGC\,5631, PGC\,28887, UGC\,3960 and
those being SRs at $R_e/2$ and FRs at $R_e$ are IC\,719, NGC\,770, NGC\,3073, NGC\,3757, NGC\,4259, 
NGC\,4803, NGC\,7710. All except two of these are either near the dividing line,
or specific cases such as, again, NGC\,4550-like systems. 
The two discrepant cases, namely NGC\,4476 and PGC\,28887, have central decoupled
kinematic components with a radial size larger than $R_e/2$.
The fact that 2$\sigma$ galaxies change class and that we still have some of
them above the threshold even within $R_e/2$ (NGC\,448 and NGC\,4473)
is expected, as the corresponding counter-rotating components span a range of 
spatial extent and luminosity contribution which directly affect the ${\lambda_R}$ 
measurements: 2$\sigma$ galaxies which are Slow Rotators have a high enough contribution within 
the considered aperture to significantly influence the measured stellar angular momentum.

As mentioned, 30 out of 36 SRs at $R_e$ are also SRs at $R_e/2$: for 22 of
these, the \Sauron\ data does not reach 1~$R_e$. Considering a
simple extrapolation of their $\lambda_R$ profiles (see also E+07), there is little chance that
any of these cross the threshold between SRs and FRs at $R_e$. 
We therefore advocate the use of one effective radius $R_e$ as the main scale to probe FRs and SRs, 
considering that doing so focuses the classification on a central but fixed and significant fraction (50\%) 
of its luminosity. 

By defining our classification criteria to such central regions,
we may be weighting more towards dissipative processes, and
consequently avoiding regions dominated by the dry assembly of galaxies \citep{Khochfar+03} 
which is thought to mostly affect the outer parts 
\citep{Naab+07, Naab+09, Hopkins+09b, Hoffman+10, Bois10, Oser+10}. We may also consequently miss very large KDCs
\citep{Coccato+09} or e.g., any outer signature of interaction.
This should obviously be kept in mind when interpreting the results: depending
on the formation and assembly history, we expect galaxies to have significantly
different radial distribution of their stellar angular momentum. 

The use of an aperture radius of 1~$R_e$ is motivated by the following facts: 
it has discriminating power, as shown in the previous Section, and we 
expect differences to be apparent in simulations (see Sect.~\ref{sec:slow}); such a scale is 
accessible to integral-field spectroscopy as well as modern numerical simulations; it
is large enough that it should not significantly suffer from variations in the 
observational conditions (e.g., seeing); it traces a significant fraction of
the stellar mass, namely about 50\% for systems with shallow spatial variations
of their stellar populations. 
We note here that we obtain consistent results with a smaller aperture, $R_e/2$, 
besides the change of class for a few flattened SRs from the 2$\sigma$ kinematic group.
We acknowledge, however, that this consistency may not hold for arbitrarily
large apertures, even though the criterion itself is a function of the aperture size.

Our refined criterion is motivated to respect the relation between dynamical
structure and apparent shape, which should increase its robustness to e.g.,
changing rotation at different radii. We also note that the criterion itself is
empirically determined, and changes depending on the scale used. One should
avoid using the distribution measured at e.g., $R_e$, to classify galaxies based
on data from grossly different spatial scales. It is also critical for any comparison 
of simulations with observed galaxies to measure these parameters consistently, 
using the same spatial extent.
\begin{figure}
\centering
\epsfig{file=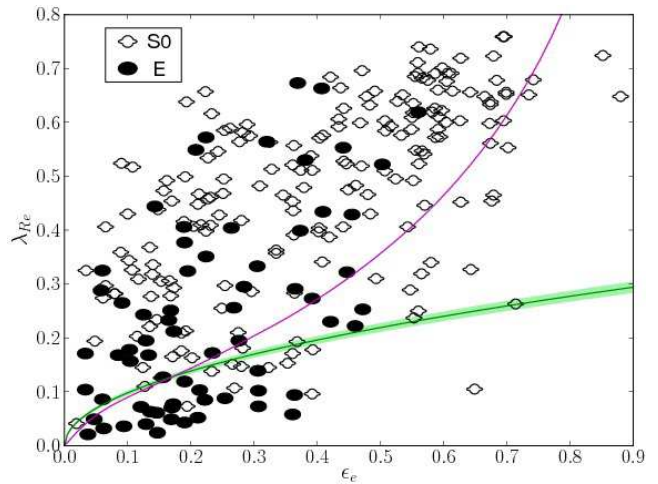, width=\columnwidth}
\caption{$\lambda_R$ versus the ellipticity $\epsilon$ 
within an aperture of $R_e$. The lines are as in Fig.~\ref{fig:LRLR2_Flag}. The filled ellipses and open symbols
are galaxies with a morphological type $T < -3.5$ (E's, ellipticals) and $T \geq -3.5$
(S0, lenticulars), respectively. Note the many E's which are Fast Rotators
(above the green line).}
\label{fig:Types}
\end{figure}

\subsection{Robustness of the new classification scheme}

The new proxy for Slow and Fast Rotators (SR and FR, respectively) 
differs from the previous constant threshold criterion in two main ways. Firstly, at high ellipticities it
reaches higher $\lambda_R$ values ($\sim 0.25$ for $\epsilon = 0.8$). Secondly, the new relation goes to zero
for very low ellipticities (apparently round galaxies). 
The three non-rotators (red circles in Fig.~\ref{fig:LRVS_NDR})
which are close to the defined relation are all very probably nearly face-on rapidly rotating
galaxies which would be significantly above the line if they were viewed edge-on.
For a Fast Rotator to be consistent with no rotation requires very low inclination and therefore
extremely round isophotes \citep[see e.g.][]{Jesseit+09}. 
The new relation works significantly
better at disentangling such cases from the truly low angular momentum galaxies.
This partly comes from the fact that the dependency on the ellipticity
(criterion $\propto \sqrt{\epsilon}$) somewhat mimics the variation of
$\lambda_R$ and $\epsilon$ due to inclination effects. 

The new criterion defined should miminise contamination and mis-classification,
but as for any empirically designed classification, we expect 
some ambiguous cases, or systems for which it is hard to conclude.
There are, for example, two galaxies, namely NGC\,5173 and NGC\,3757, which coincidentally have 
the same ${\lambda_R}_{e/2}$ and ${\epsilon}_{e/2}$, 
and lie at the very limit between SRs and FRs (NGC\,3757 is in fact a galaxy with a bar which perturbs the
ellipticity measurement). Three non-regular rotators are significantly above the curve (to
be compared with the total of 224 Fast Rotators), namely NGC\,770, NGC\,5485, NGC\,7465: NGC\,770 is a
galaxy with a known counter-rotating disc \citep{Geha+05}, NGC\,5485 is one of
the rare galaxies with prolate kinematics (as NGC\,4621), and NGC\,7465, which is the non-regular rotator with
the highest ${\lambda_R}_e$ value, is an interacting system forming a pair with NGC\,7464 \citep{LiSeaquist94} and
shows a complex stellar velocity field with a misaligned central disc-like component.

The probability of a galaxy to be misclassified as a Slow (or Fast) Rotator is 
hard to assess. We can at least estimate the uncertainty on the number of Slow
Rotators in our sample by using the uncertainty on the measurements 
themselves ($\lambda_R$ and $\epsilon$), the observed distribution of
points, and the intrinsic uncertainty in defining the threshold for
${\lambda_R^N} = {\lambda_R} / \sqrt{\epsilon}$. 
Using $R_e$ as the reference aperture, we estimate the
potential contamination of SRs by FRs by running Monte Carlo
simulations on our sample (assuming gaussian distribution for the uncertainty on 
$\epsilon$ and $\lambda_R$ of 0.05) to be $\pm 6$ galaxies ($2 \sigma$). 
We obtain a relative fraction of $\sim 14 \pm 2$\% of SRs in the full
\atlas\ sample of ETGs, which represents 4\% of the full parent sample of 871
galaxies (Paper~I). This is much lower than the 25\% quoted in E+07,
but as mentioned above, this is due to the flatness of the luminosity
distribution of the original \Sauron\ sample.

\subsection{Slow, Fast Rotators, and Hubble types}
\label{sec:Hubble}

We now examine the Hubble type classification in the light of our new scheme 
to separate Fast and Slow Rotators. 
In Fig.~\ref{fig:Types}, we show the distribution of galaxies in the
$\lambda_R$-$\epsilon$ diagram using the two main classes of
``Ellipticals'' or E's ($T < -3.5$) and ``Lenticulars'' or S0's 
\citep[$T \geq -3.5$, as defined in][]{Paturel+03}. 

The \atlas\ sample of 260 ETGs includes 192 S0's and 68 E's. As expected,
E's in the \atlas\ sample tend on average to be more massive and rounder than S0's.
We therefore naturally retrieve the trend that E's tend to populate the left
part of the diagram, and within the SR class, there is a clear
correlation between the apparent ellipticity and being classified as an E or S0,
the latter being all more flattened than $\epsilon = 0.2$.
Ellipticals also tend to be in the lower part of the diagram
(low value of $\lambda_R$), while the highest
$\lambda_R$ values correspond to S0 galaxies. Most Slow Rotators which are not
2$\sigma$ galaxies are classified as E's (23/32). 

As expected, the vast majority of galaxies with ellipticities ${\epsilon}_{e} > 0.5$ are S0s. 
However, 20\% of all Fast Rotators (45/224) are E's, and 66\% of all E's in the
\atlas\ sample are Fast Rotators. 
Also the fact that all 2$\sigma$ galaxies except one (NGC\,4473) are classified as S0's
demonstrate that global morphology alone is not sufficient to reveal the dynamical state of ETGs.
The E/S0 classification alone is obviously not a robust way to assess the dynamical state of a galaxy. 
There is in fact no clear correlation between ${\lambda_R} / \sqrt{\epsilon}$ and the morphological type $T$, 
besides the trends mentioned here. 
From Fig.~\ref{fig:Types}, we expect a significant fraction of galaxies classified
as E's to be inclined versions of systems which would be classified as S0s when
edge-on, and just separating ETGs into Es and S0s is therefore misleading.

\subsection{Properties of Slow and Fast Rotators}
\label{sec:propFS}
\begin{figure}
\centering
\epsfig{file=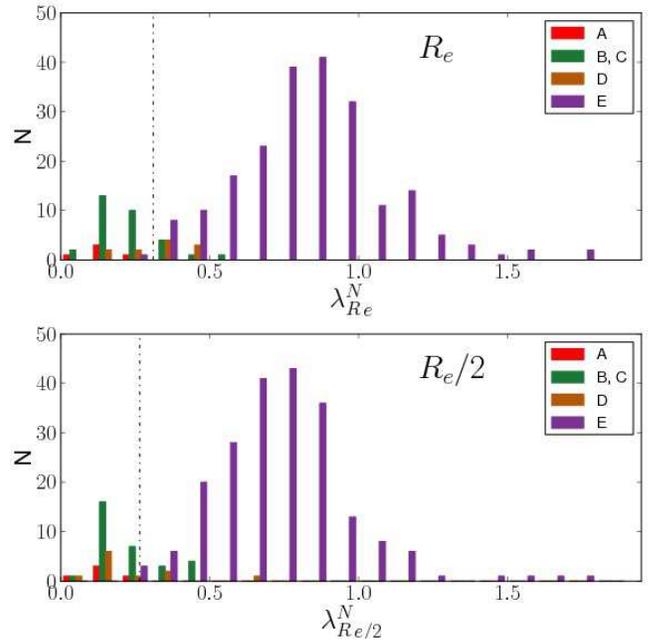, width=\columnwidth}
\caption{Histograms of galaxies in the \atlas\ sample showing the distribution
of ${\lambda_R^N} = {\lambda_R} / \sqrt{\epsilon}$ within an aperture of $R_e
/2$ (top panel) and $R_e$ (bottom panel) for non-rotators (A), 
featureless non-regular rotators with or without a KDC (B and C) in green, 2$\sigma$ galaxies (D) 
in orange, and non-regular rotators (E) in purple. The vertical dashed lines in each
panel show the limit set (0.265 and 0.31, for $R_e/2$ and $R_e$, respectively) between Slow and Fast Rotators for
both apertures. To avoid confusion, we have excluded the two non-rotators,
NGC\,3703 and NGC\,4733, which are assumed to be face-on fast rotators.}
\label{fig:HistLStar}
\end{figure}

The detailed distribution of galaxies in {\em groups a, b, c, d and e}
as defined in Paper~II is shown in Fig.~\ref{fig:HistLStar} using
histograms of ${\lambda_R^N} = {\lambda_R} / \sqrt{\epsilon}$ values both for
$R_e$ and $R_e/2$. The {\em group e} galaxies, and
consequently the Fast Rotators, peak at a value of around 0.75 within $R_e/2$
and 0.85 for $R_e$. Slow Rotators are defined as galaxies with ${\lambda_R^N} < k_{FS}$, mostly associated 
with galaxies from {\em groups a to d}, which represent the lower tail of that distribution with
some small overlap with the {\em group e}.  We provide the \Sauron\ stellar velocity and velocity
dispersion maps of all 36 Slow Rotators in Fig.~\ref{fig:SlowVmaps} and
\ref{fig:SlowSmaps} of Appendix~A. We refer the reader to \cite{Krajnovic+11}
for all other \Sauron\ stellar velocity maps.

\begin{figure}
\centering
\epsfig{file=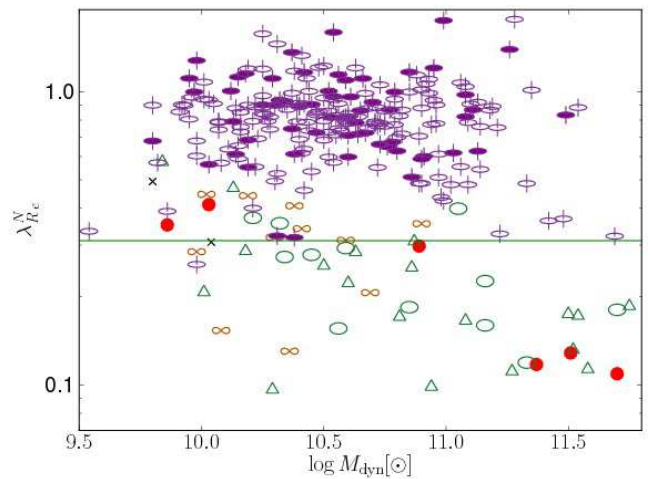, width=\columnwidth}
\caption{${\lambda_R^N}_{e}$ (log-scale) versus dynamical masses $M_{\rm dyn}$. Symbols are
as in Fig.~\ref{fig:LRLR2_Flag}. The horizontal green solid line indicates the limit
between Fast and Slow Rotators.}
\label{fig:LRNMass}
\end{figure}
\begin{figure}
\centering
\epsfig{file=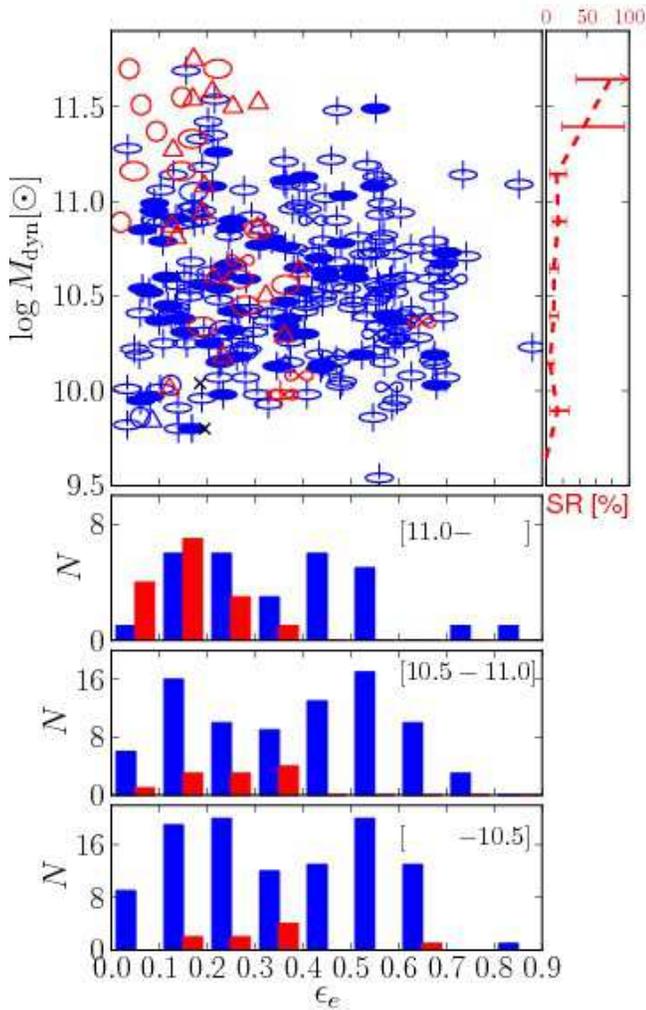, width=\columnwidth}
\caption{Dynamical masses $M_{\rm dyn}$ versus ${\epsilon}_{e}$. The top left panel
shows the distribution of \atlas\ galaxies with blue and red colours now correspond to
Fast and Slow Rotators, respectively. Symbols are as in Fig.~\ref{fig:LRLR2_Flag}.
The thick red dashed curve with errorbars on the top right panel shows the fraction of Slow Rotators 
(for 1~$R_e$) for mass bins with widths of 0.25 in $\log{M_{\rm dyn}}$ with the corresponding labels given 
on top x-axis.
The three lower panels present the
ellipticity within $R_e$ histograms in bins of masses ($\log{M_{\rm dyn}}$), indicated in brackets
in each sup-panel. Note the increased fraction of Slow Rotators in the 2
highest mass bins, and the shift of the ellipticity of Slow Rotators above and
below $10^{11}$~M$_{\odot}$.}
\label{fig:EpsMass}
\end{figure}

We re-emphasise in Figs.~\ref{fig:LRNMass} and \ref{fig:EpsMass} the trend for Slow Rotators to
be on the high mass end of our sample. 
Slow Rotators span the full range of
dynamical masses present in the \atlas\ sample. However, most non-rotators and
galaxies with KDCs have masses above $10^{10.75}$~M$_{\odot}$. If we exclude
the three potential face-on Fast Rotators (see Sect.~\ref{sec:LambdaR}), 
all non-rotators have masses above $10^{11.25}$~M$_{\odot}$.
These non-rotators and KDC galaxies clearly have a different mass distribution
than 2$\sigma$ galaxies which are all, except NGC\,4473, below $10^{10.75}$~M$_{\odot}$.
The normalised ${\lambda_R^N}_e$ value for
Slow Rotators tend to decrease with increasing mass (Fig.~\ref{fig:LRNMass}): the mean
${\lambda_R^N}_e$ values for Slow Rotators below and above a mass of $10^{11.25}$~M$_{\odot}$
are about 0.22 and 0.13, respectively. Fast Rotators overall seem to
be spread over all ${\lambda_R^N}_e$ values up to a dynamical mass $10^{11.25}$~M$_{\odot}$ where
we observe the most extreme instances of Slow Rotators (e.g., non-rotators).
In Fig.~\ref{fig:EpsMass} we show ${\epsilon}_{e}$
with respect to the dynamical mass $M_{\rm dyn}$ where we have
coloured each symbol following the Fast (blue) and Slow (red) Rotator classes. 
Fig.~\ref{fig:EpsMass} also shows the fraction of Slow Rotators with respect to
the total number of galaxies within certain mass bins: Slow Rotators represent
between 5 and 15\% of all galaxies between $10^{10}$ and $10^{11.25}$~M$_{\odot}$. 
As already mentioned, above $10^{11.25}$~M$_{\odot}$ the fraction of Slow
Rotators shoots up very significantly, reaching 45 and 77\% in the last two mass
bins below and above $10^{11.5}$~M$_{\odot}$, respectively.

There is a tendency for Slow Rotators above $10^{11}$~M$_{\odot}$ to
have rounder isophotes with ellipticities $\epsilon_{e}$ between 0 and 0.3 and
mostly below 0.2, while most Slow Rotators below $10^{11}$~M$_{\odot}$ have
ellipticities distributed between 0.2 and 0.4 (Fig.~\ref{fig:EpsMass}).
The weak trend for Slow rotators at higher mass to have lower ${\lambda_R^N}_{e}$
could thus be associated with the corresponding ellipticity decrease.
Fast Rotators seem to be smoothly distributed over the full range of
ellipticities between $10^{9.75}$~M$_{\odot}$, near the lower mass (luminosity) cut 
of our sample, and $10^{11.25}$~M$_{\odot}$ above which Slow Rotators clearly dominate in numbers.
All these results are also valid when using a smaller aperture ($R_e/2$).

\subsection{Isophote shapes and central light profiles}
\begin{figure}
\centering
\epsfig{file=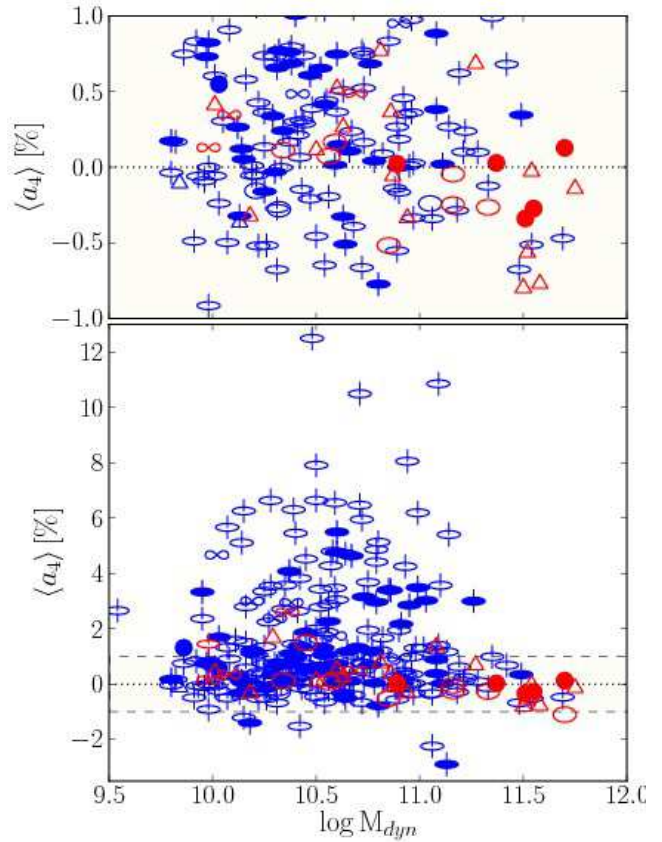, width=\columnwidth}
\caption{The averaged $a_4$ (in \% within 1~$R_e$) versus the dynamical mass for
galaxies in the \atlas\ sample. 
The top panel is a zoomed version of the bottom panel, including only $a_4$
values between $-1$ and $+1$. Symbols are as in the top panel of Fig.~\ref{fig:EpsMass}. }
\label{fig:Massa4}
\end{figure}
\begin{figure}
\centering
\epsfig{file=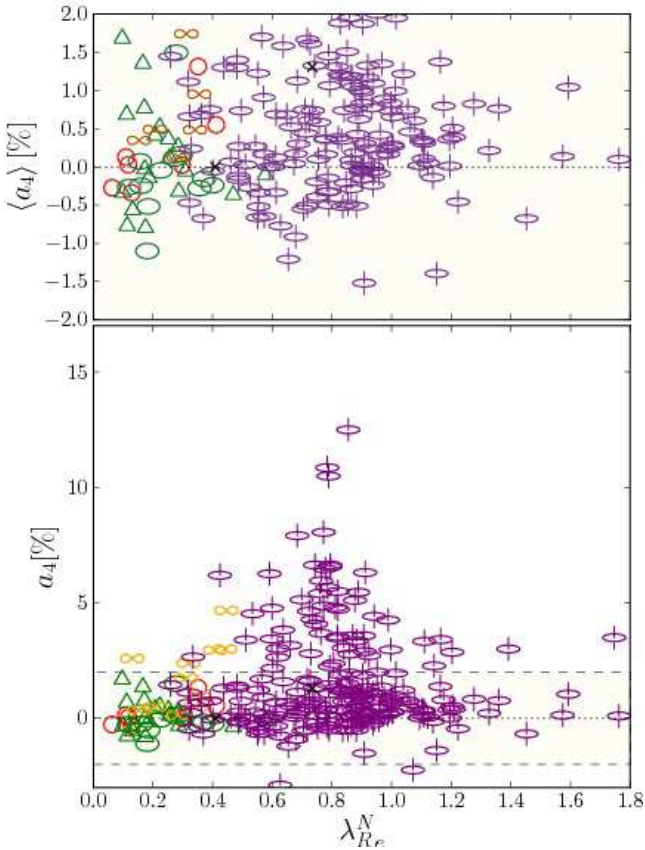, width=\columnwidth}
\caption{${\lambda_R^N}_{e}$ versus the averaged $a_4$ (in \% within 1~$R_e$). 
The top panel is a zoomed version of the bottom panel, including only $a_4$
values between $-1$ and $+1$. Symbols are as in Fig.~\ref{fig:LRLR2_Flag}. 
The vertical green light represents the threshold between Slow and Fast Rotators, as
defined in the present paper.}
\label{fig:LRa4}
\end{figure}
\begin{figure}
\centering
\epsfig{file=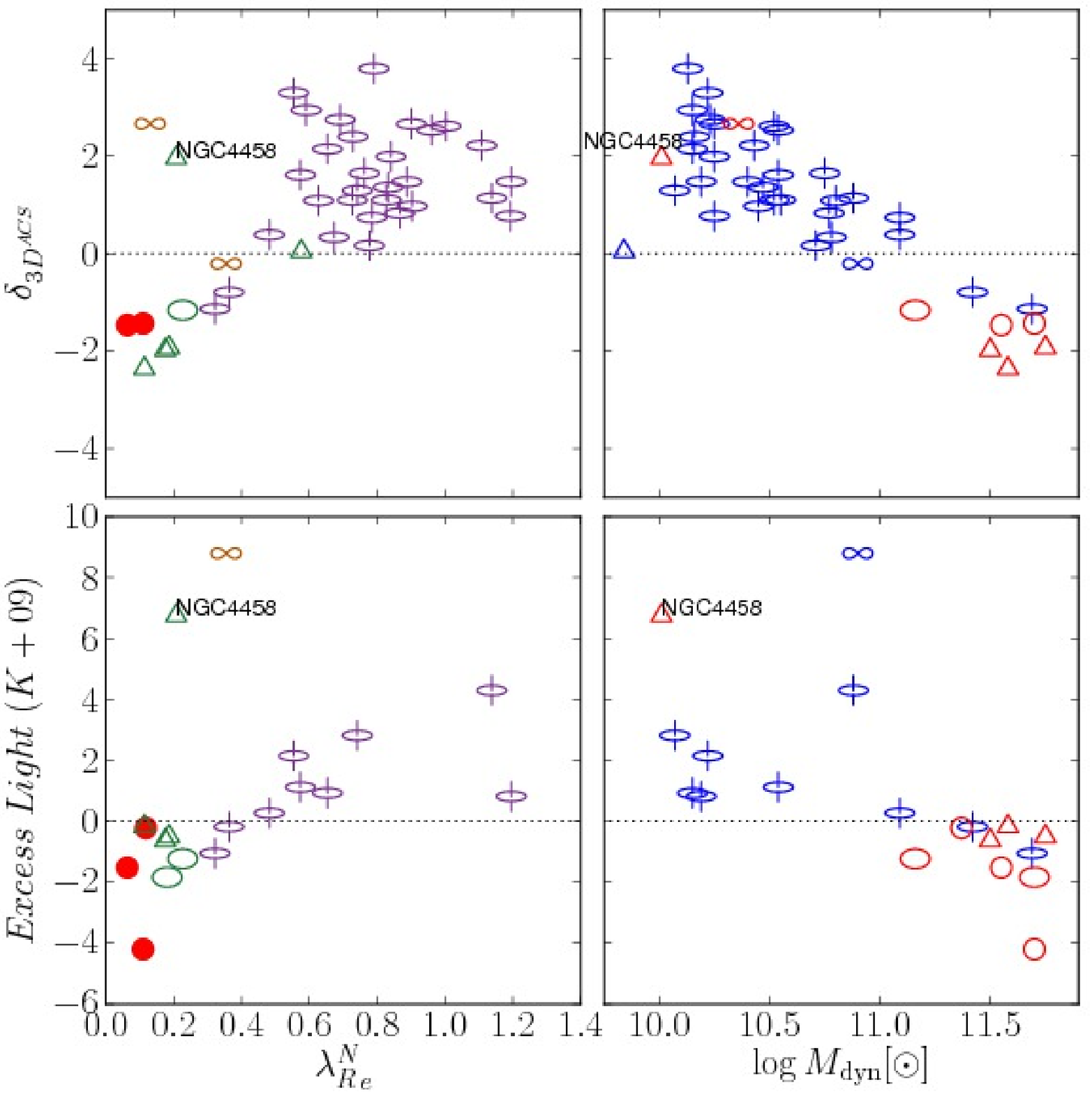, width=\columnwidth}
\caption{The central light deficit as derived by
\citet[][top panel]{Glass+10} and \citet[][bottom panel]{Kormendy+09}, for galaxies in common with
the \atlas\ sample, versus ${\lambda_R^N}_e$ (left panels) and the dynamical mass $M_{\rm dyn}$ (in log, right panels). 
Symbols in the right panels are as in Fig.~\ref{fig:LRLR2_Flag}, while
in the right panels, Slow Rotators are indicated as red symbols, and Fast Rotators as blue symbols. 
NGC\,4458 is emphasised as only galaxy with a KDC (triangle) to have a positive
value for the light excess.}
\label{fig:ExcessMass}
\end{figure}

Massive Slow Rotators have, as expected, slightly boxy isophotes, as shown
in Figs.~\ref{fig:Massa4} where we present $\left< a_4 \right>$, the luminosity weighted average of $a_4$
within 1~$R_e$, versus the dynamical mass. This is a
known result, specifically emphasised by \cite{KB96} who suggested the use of
the $a_4$ parameter, quantifying the degree of boxiness or disciness of the
isophotes, to assess the dynamical status of early-type galaxies. 
All 9 galaxies within the \atlas\ sample with masses larger than $10^{11.5}$~M$_{\odot}$
have $a_4$ values which are negative or very close to zero, but these include 2 fast
rotators, namely NGC\,3665 and NGC\,4649. In fact 70\% of the Slow Rotators more massive 
than $10^{11}$~M$_{\odot}$ are boxy, while in the same mass range only 30\% of the Fast Rotators are.
Below a mass of $10^{11}$~M$_{\odot}$, we observe the same overall fraction
($\sim 25$\%) of boxy systems in both Slow and Fast Rotators.
This means that the relative fraction of Fast Rotators which are boxy is nearly constant
with mass, while there a drastic change of the boxiness/disciness in the
population of Slow Rotators around a mass of $10^{11}$~M$_{\odot}$.
We also note that most discy Slow Rotators exhibit a KDC,
or are 2$\sigma$ galaxies. Since all 2$\sigma$ galaxies are discy and 
the identification of such systems depends on the viewing angle, we should
expect that some of these discy Slow Rotators are bona fide 2$\sigma$:
a confirmation of this hypothesis requires detailed 
modelling of the photometry and stellar kinematics.

In Fig.~\ref{fig:LRa4}, we now show how $\left< a_4 \right>$ varies with the normalised value $\lambda_R^N$ with
the symbols of the different kinemetry groups. Non regular rotators with no specific kinemetric feature
({\em group b}) are more often boxy. Galaxies with KDCs ({\em group c}) can be both discy or boxy.
Larger positive disciness values are reached for higher $\lambda_R$ values 
as already emphasised in E+07, and galaxies with $\left< a_4 \right>$
larger than 3\% are all Fast Rotators. A little more than 20\% of all Fast
Rotators are boxy (48/224), but most of them (32) with rather low absolute values
(average boxiness of less than 0.5\%), and a few (e.g., NGC\,3489, NGC\,4233) 
because of the impact of dust on the isophote shapes. 
Among these boxy Fast Rotators, only 17\% are clearly barred (8/48): 
considering the size of that sub-sample, this is not significantly 
different from the 28\% of Fast Rotators which are clearly barred (with this fraction
of barred galaxies to be considered as a lower limit). 
Apart from the mass trend mentioned above, there therefore seems to be no 
simple link between $a_4$ of galaxies in the \atlas\ sample and the Slow and
Fast Rotator classes.

We also examine whether there is an existing link between the 
apparent stellar angular momentum measured via $\lambda_R$ within 1~$R_e$
and the central light excess (or deficit) : these central departures from 
simple photometric profiles (Sersic laws) have been interpreted in various contexts but the main processes
which have been called upon are dissipational processes (gas filling in
the central region and forming stars) to explain the light excesses, and black
hole scouring (ejection of stars by binary black holes).
In this context, \cite{Kormendy+09} have
recently proposed that it represents an important tracer of the past history of the
galaxy, and suggested the existence of a dichotomy within the E (elliptical) class of galaxies
\citep[see also papers by][]{Khochfar+05,Hopkins+09a, Hopkins+09b, Hopkins+09c}.
In Fig.~\ref{fig:ExcessMass} we present the central excess light value 
obtained by \citet[][bottom panel]{Kormendy+09} and \citet[][top panel]{Glass+10} 
in terms of the dynamical mass ${\lambda_R^N}_e$ (left panels) and $M_{\rm dyn}$
(right panels). 
We recover the trend already mentioned in \cite{Cote+07}, \cite{Kormendy+09} and
in \cite{Glass+10}, that central light
excesses correlate with luminosity (or mass).
There is a clear trend for Fast Rotators to have central light excesses, and Slow Rotators to exhibit
light deficits. The central light excess is, however, not strongly correlated with the
distance to the threshold defining Slow and Fast Rotators, as e.g., Fast Rotators
span a wide range of $\lambda_R / \sqrt{\epsilon}$ values irrespective of the
measured central light excess. 
There is a remarkable exception in Fig.~\ref{fig:ExcessMass} for 2$\sigma$ galaxies which are Slow
Rotators but have light excesses: there, morphology and photometry alone 
fail to reveal the nature of the underlying stellar system, although there may
exist associated photometric signatures. NGC\,4458 also stands out as a Slow Rotator, with
a KDC ({\em group c}) and having a light excess which seems to correspond to a
very central stellar disc \citep{Morelli+04,Ferrarese+06}. 
The existing trend between $\lambda_R$ and total luminosity or mass mentioned in the present 
paper may be enough to explain the observed trend, e.g., that Slow Rotators exhibit light {\em
deficits}. However, the small number of objects in Fig.~\ref{fig:ExcessMass} is obviously a major concern
in this context, and we should wait for access to larger datasets before we can draw firm conclusions.

\section{Discussion}
\label{sec:discussion}
The results presented in the previous Sections provide an updated view at 
ETGs in the nearby Universe, which we first briefly summarise now, before
we further discuss the two {\em families} of Fast and
Slow Rotators in turn, and mention how these families can relate to
standard physical processes often invoked in the context of galaxy formation and
assembly.

The vast majority of ETGs in our sample are Fast Rotators: they dominate
in numbers and represent nearly 70\% of the stellar mass in ETGs. 
Fast Rotator are mostly discy galaxies spanning a wide range of 
apparent ellipticities, the most flattened systems having ellipticities
consistent with the ones of spiral galaxies. 
Slow Rotators represent only 15\% of all ETGs in the \atlas\ sample, and
only dominate the high mass end of the ETG distribution.
We witness three main types of Slow Rotators out of the 36 present in the \atlas\ sample: 
\begin{itemize}
   \item Non-rotating galaxies (4, excluding NGC\,6703 which may be a face-on Fast Rotator), 
      which are all more massive than $10^{11.25}$~M$_{\odot}$, and appear
      round, namely NGC\,4374, NGC\,4486, NGC\,4436 and NGC\,5846;
   \item Galaxies which cover a large range of masses (27), often with KDCs (18/27, including
      NGC\,4476 in this set), and are never very flattened with $\epsilon_e$ smaller than 0.4. 
      This includes 7 galaxies which have KDCs observed as counter-rotating structures;
   \item Lower-mass flattened 2$\sigma$ galaxies (4), which are interpreted
as including two large-scale stellar disc-like counter-rotating components.
\end{itemize}
The \atlas\ survey probes various galaxy environments, and 
obviously, the fraction of observed Fast and Slow Rotators 
would change with samples of galaxies biased towards e.g.,
higher galaxy densities (Paper~VII).

\subsection{Fast rotators}
\label{sec:MC}

We can now use our knowledge of the photometric and kinematic structures of Fast Rotators to
predict roughly how a given sample of galaxies with particular properties would 
be distributed if viewed at random inclinations on the sky. A simple
prescription could be applied to the family of Fast Rotators, 
as the vast majority of these exhibits regular stellar velocity maps consistent with apparent disc-like rotation
with no significant misalignments between the photometric and kinematic axis
(see Paper~II for more details on this specific issue). If
the galaxies illustrated in Fig.~\ref{fig:LREpsMass}, selected to be close to the $\beta
= 0.65 \times \epsilon$ relation for oblate systems, are indeed viewed near edge-on, 
then Fast Rotators span quite a broad range of intrinsic flattening within 1~$R_e$. 
In Fig.~\ref{fig:LREpsMass}, the dashed lines emphasise
the effect of inclination for galaxies along this relation (see also Appendix~B). 

\begin{figure}
\centering
\epsfig{file=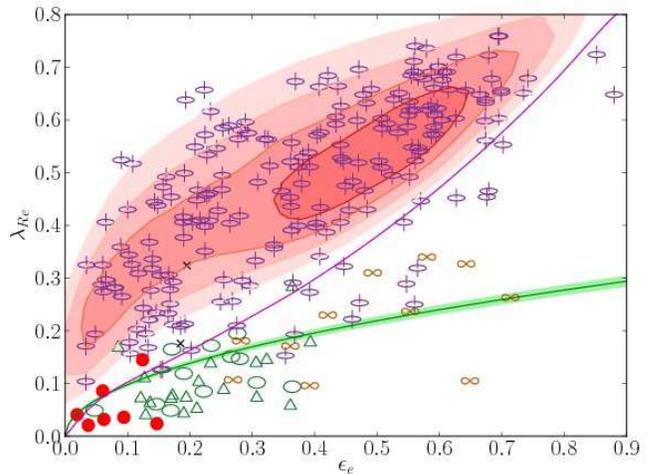, width=\columnwidth} 
\caption{${\lambda_R}_e$ versus apparent ellipticity $\epsilon_e$ with flags
indicating the observed kinematic structure with symbols as in Fig.~\ref{fig:LRLR2_Flag}. The coloured
contours show the result of Monte Carlo realisations as described in the text.
The green line shows the limit between Slow and Fast Rotators.}
\label{fig:LREps_MC}
\end{figure}

We therefore performed Monte-Carlo simulations of a large sample
of galaxies following a prescription similar to the one from C+07 (see
also Appendix C). We take the distribution for 
the intrinsic ellipticities $\epsilon_{intr}$ of the simulated sample
as a Gaussian centred at $\epsilon_{0} = 0.7$ with a width $\sigma_{\epsilon} = 0.2$ for an aperture
of 1~$R_e$. We fixed the $\beta$ anisotropy
distribution also as a Gaussian with a mean of $m_{\beta} = 0.5$ and a dispersion of $\sigma_{\beta}
= 0.1$ truncated at $0.8 \times \epsilon_{intr}$ (see C+07).
The result of the simulation (50,000 galaxies) is shown in Fig.~\ref{fig:LREps_MC}, and
it is qualitatively consistent with the distribution of Fast Rotators (see also
a similar simulation, but for $R_e/2$ in Appendix~B, Fig.~\ref{fig:LR2Eps2_MC}).
This tells us that in this context a reasonable approximation for such galaxies is a
set of oblate systems with ellipticities peaking around 0.7, with most of the
objects between 0.4 and 0.8. This confirms the visual impression provided in Fig.~\ref{fig:LREpsMass}
illustrating the effect of the inclination (dashed lines).
This Monte Carlo simulation also tells us that we should expect, from a sample
of 224 Fast Rotators, an average of $\sim 4\pm2$ galaxies with $\lambda_R$ below
0.1, which is consistent with our observed sample.

\subsection{Slow rotators}

Fig.~\ref{fig:LREps_MC} shows, as expected, that slow rotators are clearly inconsistent
with the previously simulated dataset. This confirms that the central regions of
Slow Rotators are not similar to oblate systems assuming some simple anisotropy-flattening
relation as done in the previous Section. Even an ellipticity distribution different from
what was assumed in this simulation would not help to cover the region 
where Slow Rotators are located in a $\lambda_R$-$\epsilon$ diagram. Note that galaxies in {\em group d}
have morphologies consistent with oblate systems but have unusually strong
tangential anisotropies associated with the presence of two counter-rotating
stellar discs. Observed misalignments between the kinematic and photometric axis
in Slow Rotators also argue for these to be a different family of
galaxies (Paper~II) including non-axisymmetric and/or mildly triaxial systems.

As demonstrated \cite{Jesseit+09} and confirmed in Paper~V
via numerical simulations of mergers, triaxial remnants which are Slow Rotators 
when viewed edge-on tend to appear as Slow Rotators (stay below the limiting line)
for any projection. In the previous Section, we have seen that there is only a
low probability for a Fast Rotator to be viewed (due to inclination effects) as a Slow Rotator
\citep[something also emphasised by][]{Jesseit+09}, and here we suggest that
most galaxies appearing as Slow Rotators would still be Slow Rotators if viewed edge-on.
The only cases which are expected to potentially exhibit ambiguous classifications
are prolate objects \citep{NaabBurkert03,Jesseit+05}. 
We have two clear cases in our \atlas\ sample: NGC\,4261, and NGC\,5485, the
latter being classified as a Fast Rotator here. NGC\,5485 is a peculiar
galaxy with its photometric major-axis being the symmetry axis for the dust lane
and stellar rotation.

\subsection{Early-type galaxy formation and assembly processes}
\label{sec:slow}
\begin{figure}
\centering
\epsfig{file=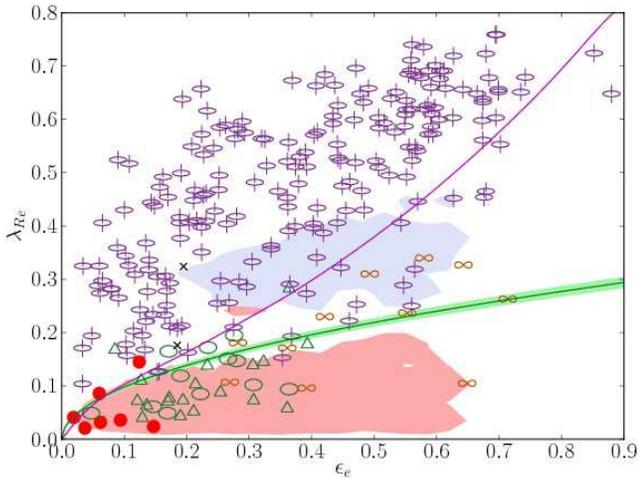, width=\columnwidth}
\caption{${\lambda_R}_e$ versus apparent ellipticity $\epsilon_e$ within 1~$R_e$ with
symbols as in the bottom panel of Fig.~\ref{fig:LREps_MC}.
The blue and red filled coloured areas show the region where 1:1 and 2:1 merger remnants lie
from the study of \citet{Bois11}: the red area corresponds to the merger remnants with
a retrograde main progenitor (w.r.t. the orbital angular momentum) while the
blue area corresponds to a prograde main progenitor.
The green lines show the limit between Slow and Fast Rotators
and the magenta line is as in Fig.~\ref{fig:LRLR2_Flag}.}
\label{fig:LREps_Mergers}
\end{figure}

Various physical processes often invoked for the formation and assembly of galaxies can
contribute to or influence the specific angular momentum of the central stellar component
of early-type galaxies, and consequently the measured $\lambda_R$ value, and we
briefly review these here in the context of the classification of ETGs into Fast and Slow Rotators.

Dissipative processes followed by star formation should generally help
preserving (or rebuiding) stellar rotation in galaxies \citep[see e.g., an early
discussion in the context of early-type galaxies in][]{Bender+92}.
Accretion of gas from external sources, either via the large-scale filaments
\citep[e.g.][]{Sancisi+08,Dekel+09, KhochfarSilk09} 
or extracted by tidal forces from a gas-rich galaxy passing by, 
thus contribute to an increase of $\lambda_R$ if the gas is 
co-rotating with the main existing stellar component, assuming
that this additional gas forms stars. 
This could also be a way for galaxies to rebuild a fast rotating disc-like component.
Such a process should preferentially occur in gas-rich environments, e.g., at high redshifts ($z$ larger than about 2) 
and/or in low-density regions, disfavouring for instance the inner regions of dense clusters at moderate to low redshifts.
An extreme version of such a mechanism is the case of very high gas fractions in disc-like objects, expected to be relevant
mostly at high redshift: in such a situation, strong instabilities lead the 
galaxy to become "clumpy", with massive gas clouds
forming stars and evolving as a $N$-body system with low $N$ \citep[see e.g.][]{Elmegreen+08,Elmegreen+09}. 
Evolving in relative isolation, such a system can end up as a disc galaxy \citep[a
Fast Rotator, e.g.][]{Bournaud+08} where the contribution of its spheroidal component varies depending 
on the exact initial conditions (gas fraction, spatial distribution, angular momentum). 

Subsequent evolution due to disc instabilities, spiral density waves and bars, 
will tend to heat the stellar component \citep{SellwoodBinney02,Debattista+06, Sales+09, Minchev+10},
and decrease $\lambda_R$ accordingly, but the global stellar angular momentum of Fast Rotators 
should not change dramatically because of such secular evolution processes. 
Such perturbations could, however, be a trigger for inner gas fuelling then 
leading to the formation of a central rapidly rotating stellar
component \citep[e.g.][]{Wozniak+03}. 
Similarly, gas stripping from a galaxy, if done in a nearly adiabatic
way, should not change $\lambda_R$ too drastically, even though the system 
would morphologically and dynamically evolve on a relaxation timescale and this
may affect the morphology and dynamics of its central region.
This should also concern ram pressure stripping \citep{GunnGott72, Quilis+00,
Rasmussen+06}, or AGN feedback if the major effect remains focused
on the gas component.

For a disc-like (spiral) galaxy to become a Slow Rotator, numerical simulations
have suggested that it needs to accrete at least half of its stellar mass via
mergers \citep{Bournaud+07, Jesseit+09, Bois10, Bois11}.
As the orbital angular momentum for a (binary) merger event is often
the main contributor \citep{Khochfar+06},
major mergers can form Fast Rotators (\citet{Bois11}, hereafter Paper~VI; and see also \citet{SpringelHernquist05}) even from slowly rotating galaxy progenitors
\citep{diMatteo+09}, the outer structure being generally more significantly affected \citep{Coccato+09}. 
Numerical studies show anyway that, among binary mergers, only major 1:1 or 2:1 mergers can form
Slow Rotators, as it requires enough baryonic angular momentum to be transfered outwards
(see Paper~VI for details).
We illustrate this by indicating where the 1:1 and 2:1 major merger remnants (including
gas and star formation) conducted in Paper~VI lie in a 
$\lambda_R$-$\epsilon$ diagram in Fig.~\ref{fig:LREps_Mergers}, assuming that
the progenitors were spiral galaxies which have $\lambda_R$ values close to the
maximum value observed for our sample of early-type galaxies.  
There is a clear separation between the merger remnants which are Slow and
Fast Rotators: this corresponds to an initially different sign of the spin of the more
early-type progenitor (Paper~VI) with respect to the orbital angular
momentum as illustrated by the red and blue areas (corresponding to retrograde
and prograde spins, respectively) in Fig.~\ref{fig:LREps_Mergers}. Although such a distinction may be damped 
if we more broadly sample the input parameters for the progenitors (including their mass ratios)
or include more realistic merger trees, the criterion defined here to separate Fast 
and Slow Rotators seems to properly distinguish two families of galaxies:
the merger remnants in Paper~VI which are Fast Rotators 
all have regular velocity fields with small photometric versus kinematic misalignments, 
while most of the remnants which are Slow Rotators have kpc-size KDCs
\citep[see also][Paper~VI]{vdBosch+08, Hoffman+10}, and are spread
over a wide range of misalignment angle values.

In a similar context, ``2$\sigma$'' galaxies (10 in our sample, 4 of which are Slow Rotators)
clearly stand out in an $\epsilon$-Mass diagram within the Slow Rotator class (see Fig.~\ref{fig:EpsMass}),
and such systems could be formed when two spiral galaxies with (roughly) opposed spins merge \citep{Crocker+09}. 
Another scenario relies on the accretion of external and counter-rotating gas in
a spiral galaxy \citep{Rubin+92}. \cite{Crocker+09} recently suggested that the sense
of rotation of the remnant gas component could indicate which scenario is
preferred: associated with the thicker component for the merger scenario, or
with the thinner component for the accretion scenario. In the prototype galaxy
NGC\,4550, the gas rotates with the thicker stellar disc, favouring a merger
event. However, other 2$\sigma$ galaxies seem to show various configurations
for the gas. We should also examine more cautiously a broader range of mergers
forming such systems, as well as simulate the accretion of counter-rotating gas,
before we can go further and constrain the main formation process involved here.
In any case, such Slow Rotators very probably require an existing cold, spiral-like
progenitor and/or a gas-rich event (merger or accretion). The fact
that such Slow Rotators show positive light excess (Fig.~\ref{fig:ExcessMass})
contrary to other Slow Rotators is thus naturally explained if 
they formed and assembled via dissipative processes (accretion or merger), with the excess light being
the consequence of a secondary star formation even in the central regions \citep{Mihos+94, Hopkins+09c, Kormendy+09}.

To summarise, ways to increase the central specific angular momentum of a galaxy
are common, specially at high redshift, and often involve a gaseous dissipational process.
Among early-type galaxies, Fast Rotators may therefore be expected to 
dominate in numbers at low redshifts, as long as the observed sample is not biased
towards very dense environments (see Paper~VII).
The fastest early-type rotators should also be
linked to their gas-richer counterpart, namely the spiral galaxies. Indeed, a
galaxy like NGC4762 is an excellent illustration of a "dead" spiral with a very
high specific (stellar) angular momentum. Other Fast Rotators are the result of
a complex history which certainly includes a mix of the processes mentioned in
the present Section, and other potentially important ones such as stellar mass
loss \citep[see e.g.][]{MartigBournaud10}. 

Slow Rotators are the extreme instances 
of such a mixture of processes, where one of the most violent and disturbing mechanism, namely
major or repeated mergers, disrupted their kinematical identity and for which
there was little opportunity to transform back into a Fast Rotator: this mostly
happens either at the higher end of the mass function \citep[e.g.][]{Khochfar+06,
Oser+10}, or, specifically, for low-mass galaxies which are the result of the merging of two counter-rotating 
(gaseous and/or stellar) components. This picture is consistent with the fact
that little or no molecular gas is found in Slow Rotators \citep[][Paper~IV]{Young+11}.

The detailed structure of nearby ETGs is a consequence of such a (non-exhaustive)
list of processes: the exact distribution of ETGs in term of a specific
observable, e.g., $\lambda_R^N$ (Fig.~\ref{fig:HistLStar}), can then be simply interpreted 
as the convolution of the impact of each mechanism on that quantity  with
its relative contribution (this being obviously a naive view considering that
each process does not act independently). We should therefore naturally expect 
a broad and continuous range in the properties of nearby ETGs.
Since the relation defined to separate Fast and Slow Rotators is an 
empirical criterion, i.e. based on observable quantities such as $\lambda_R$ 
and $\epsilon$, we should also expect a continuous 
range of properties among Fast Rotators linking them to the Slow Rotators,
the final state of an individual galaxy 
depending on a given (complex) merging/accretion/evolution history. 
We indeed find galaxies near the dividing line with intermediate
properties, and there are a few misclassified systems.
Still, the defined criterion for Slow and Fast Rotators 
operates rather well when the goal is to distinguish galaxies with
complex central dynamical structures (e.g., large-scale stellar KDCs), from oblate systems with or without bars.
We therefore suggest here that Slow Rotators are the extreme instances of such a
mixture of processes where one of the most violent and disturbing mechanism, namely
major or repeated mergers, disrupted their kinematical identity and for which
there was little opportunity to transform back into a Fast Rotator. This mostly
happens either at the higher end of the mass function \citep[e.g.][]{Khochfar+06,
Oser+10}, or, specifically, for low-mass galaxies which are the result of the merging of two counter-rotating 
(gaseous and/or stellar) components. This picture is in fact consistent with the finding
that there seems to be little or no molecular gas in Slow Rotators \citep[][Paper~IV]{Young+11}.
Most Slow Rotators have a mass above $10^{10.5}$, and they clearly dominate the
ETG population above $10^{11.25}$. If the above-mentioned picture is correct,
then the smoothness of the transition between a Fast Rotator dominated 
to a Slow Rotator dominated population of ETGs should directly depend on the
availability of gas and the role of mergers. This is examined in detail in forthcoming papers, e.g., 
in \cite{Khochfar+11} where semi-analytic modelling techniques suggest that the amount of 
accreted material and the ability to cool gas and make stars play a prominent role in 
this context. More specific studies of mergers are presented in Paper~VI
(numerical simulations of binary mergers) and more specifically in a
cosmological context in \cite{Naab+11}.

\section{Conclusions}
\label{sec:conclusions}

In this paper, we have used photometric and integral-field spectroscopic data
to provide the first census of the apparent specific (stellar) angular momentum of
a complete sample of 260 early-type galaxies via the $\lambda_R$ parameter. 

We have shown that the apparent
kinematic structures can be used to define a refined and optimised criterion to disentangle
the so-called Fast and Slow Rotators. This new definition takes into account the
fact that at a similar $\lambda_R$ value, a more flattened galaxy is expected to exhibit
a much stronger anisotropy. We are using a simple proxy 
with Slow Rotators being galaxies which have a specific stellar angular momentum within $R_e$
as measured by ${\lambda_R}$ less than 0.31 (resp. 0.265) times the square root
of the ellipticity $\sqrt{\epsilon}$ measured within an aperture of 1~$R_e$ (resp. $R_e/2$).

Using this relation, we find that $86 \pm 2 $\% (224/260) of all early-type galaxies in
our \atlas\ sample are Fast Rotators. 
This result, associated with the fact that Fast Rotators
in the \atlas\ sample have aligned photometric and kinematic axes within 5\degr\
(Paper~II), suggest that Fast Rotators are simple
oblate systems (with or without bars) which span a range of intrinsic ellipticities 
between about 0.35 and 0.85. The remaining 14\% (36/260) of the \atlas\ sample
are Slow Rotators: these are distributed between the well-known massive (M$>10^{11}$M$_{\odot}$)
and rather round ($\epsilon_e < 0.4$) galaxies (4 being very massive
non-rotators), often exhibiting central Kinematically
Decoupled Components, and a set of 4 (or 11\% of all Slow Rotators) lower mass objects 
(M$_{\rm dyn} < 10^{10.5}$~M$_{\odot}$) which exhibit two large-scale 
counter-rotating stellar disc systems (see Paper~II for details).

We show that the suggested proxy, namely the $a_4$ parameter
is not efficient at distinguishing Fast Rotators from Slow Rotators. 
We also conclude that the separation of ETGs into E (elliptical) and S0 (lenticular) classes
is misleading. We do observe a trend, as expected, in the sense that most massive Slow Rotators are
classified as E's and most Fast Rotators are classified as S0's. However, 66\%
of all E's in our sample are Fast Rotators, and, apart from a maximum ellipticity
of 0.6, these are indistinguishable in our study from the rest of the 
Fast Rotator population, a result already pointed out in E+07.
We provide a quantitative and robust criterion to separate both families,
via the $\lambda_R$ parameter, while we expect only some small overlap 
around the defined threshold. 

Early-type galaxies are the end results of a complex set of formation and
assembly processes, which shaped their morphology and dynamics. 
From the fastest rotating ETGs, which look like dead and red spirals, down to
the slowest Fast Rotators, we observe no obvious discontinuity in their
integrated properties. There is, however, a clear increase in the fraction
of Slow Rotators above a mass of $\sim 10^{11.25}$~M$_{\odot}$ (see
Sect.~\ref{sec:propFS}), as well as above a certain local galaxy density (Paper~VII).
In this context, Slow Rotators represent the extreme instances
of such red sequence galaxies for which significant merging has occurred (sometimes with
gas-rich systems but then at high redshift), and being at the high end of the mass range of galaxies in
the nearby Universe, did not have the opportunity to rebuild a cold stellar
component from dissipative processes.

These results argue for a shift in the existing paradigm for early-type galaxies, 
which are generall separated in disc-like S0 galaxies and spheroidal-like E systems. 
We find that the vast majority of ETGs in the nearby Universe are Fast Rotators (with 66\% of the
E's in our sample being Fast Rotators), 
galaxies consistent with oblate systems with or without bars, 
while only a small fraction of them (less than 12\%), the
tail of that distribution encompassing the most massive objects, 
have central (mildly) triaxial structures reflecting the more standard picture of an
ellipsoidal (or spherical) stellar system.

\section*{Acknowledgements}
Part of this work has been conducted while visiting the Astronomy Department at University of Texas at Austin under the auspices of a Beatrice Tinsley Scholarship. In this context, EE would like to warmly thank Neal Evans, Gordon Orris and John Kormendy for their hospitality. EE would also like to thank John Kormendy, Laura Ferrarese and Pat C\^ot\'e for insightful discussions. EE thanks Lisa Glass and Pat C\^ot\'e for sending an ascii table with the $\delta_{3D}$ values published in \cite{Glass+10}.
This work was supported by the Agence Nationale de la Recherche under contract ANR-08-BLAN-0274-01.
This work was supported by the rolling grants `Astrophysics at Oxford' PP/E001114/1 and ST/H002456/1 
and visitors grants PPA/V/S/2002/00553, PP/E001564/1 and ST/H504862/1 from the UK Research Councils. 
RLD acknowledges travel and computer grants from Christ Church, Oxford and support from the Royal Society 
in the form of a Wolfson Merit Award 502011.K502/jd. RLD also acknowledges the support of the ESO Visitor Programme which funded a 3 month stay in 2010.
SK acknowledges support from the the Royal Society Joint Projects Grant JP0869822.
TN, SK and MBois acknowledge support from the DFG Cluster of Excellence  `Origin and Structure of the Universe'.
MC acknowledges support from a STFC Advanced Fellowship (PP/D005574/1) and a Royal Society University Research Fellowship.
MS acknowledges support from a STFC Advanced Fellowship ST/F009186/1.
NS and TD acknowledge support from an STFC studentship.
RcMd is supported by the Gemini Observatory, which is operated by the Association
of Universities for Research in Astronomy, Inc., on behalf of the
international Gemini partnership of Argentina, Australia, Brazil, Canada,
Chile, the United Kingdom, and the United States of America.
The authors acknowledge financial support from ESO.

\bibliographystyle{mn2e}
\bibliography{PaperIII_ATLAS3D_Final.bbl}

\appendix
%
%
\section{Stellar velocity and dispersion maps of Slow Rotators in the \atlas\ sample}
\label{App:vmaps}

\begin{figure*}
\centering
\epsfig{file=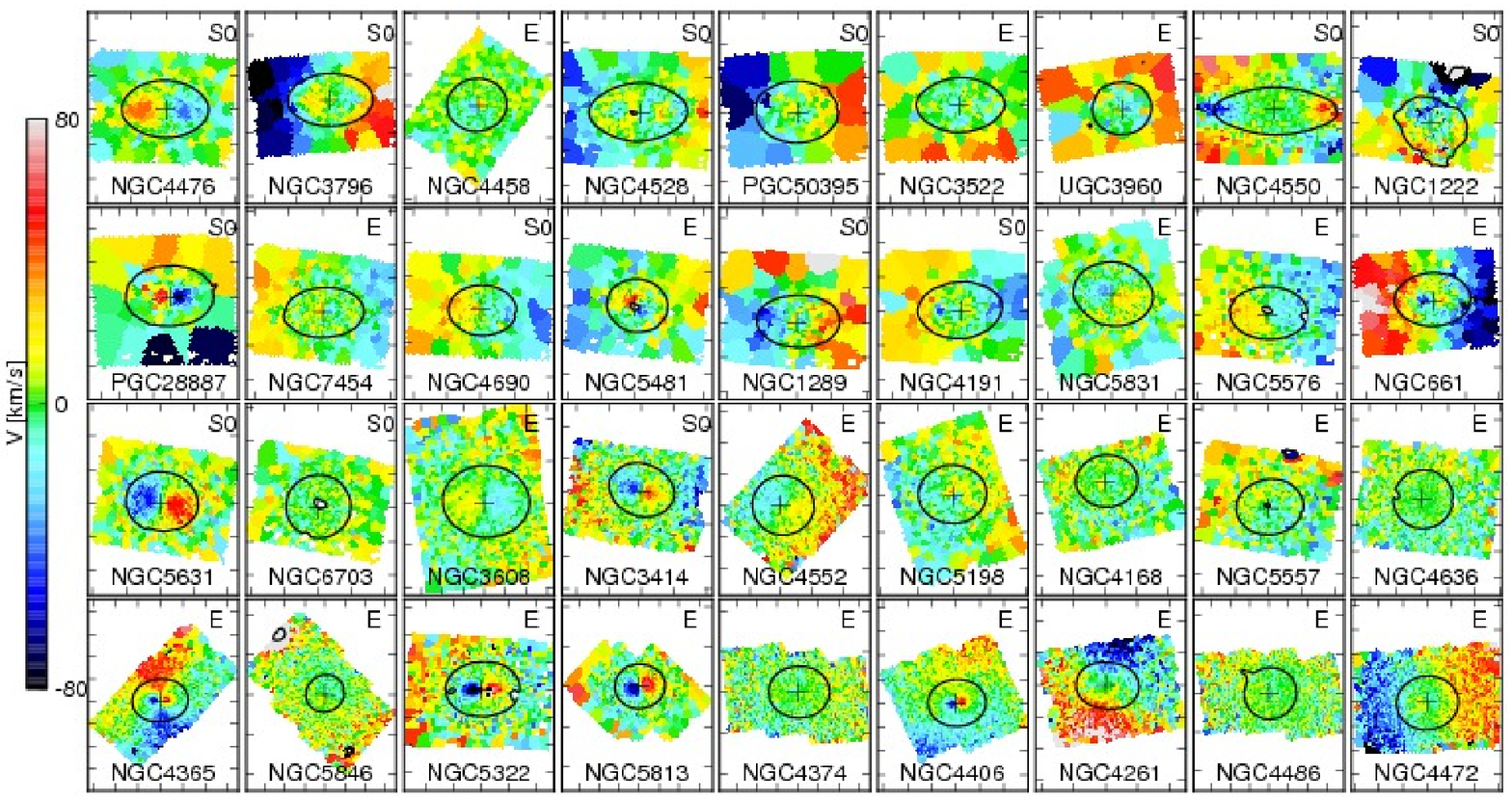, width=\hsize}
\caption{Stellar velocity fields for the 36 early-type slow rotators in the
\atlas\ sample. We used the outer photometric axis to align all galaxies horizontally.
Colour cuts have been adapted to each individual map. The solid black contour
correspond to a representative isophote and the centre of the galaxy is indicated by a cross.
From top to bottom, left to right, the order follows the (increasing) dynamical
mass values. The names of the galaxies and their Hubble types (E or S0) are also indicated. The colour cuts have
been set up to $\pm 80$~[km.s$^{-1}$] for all maps.}
\label{fig:SlowVmaps}
\end{figure*}
\begin{figure*}
\centering
\epsfig{file=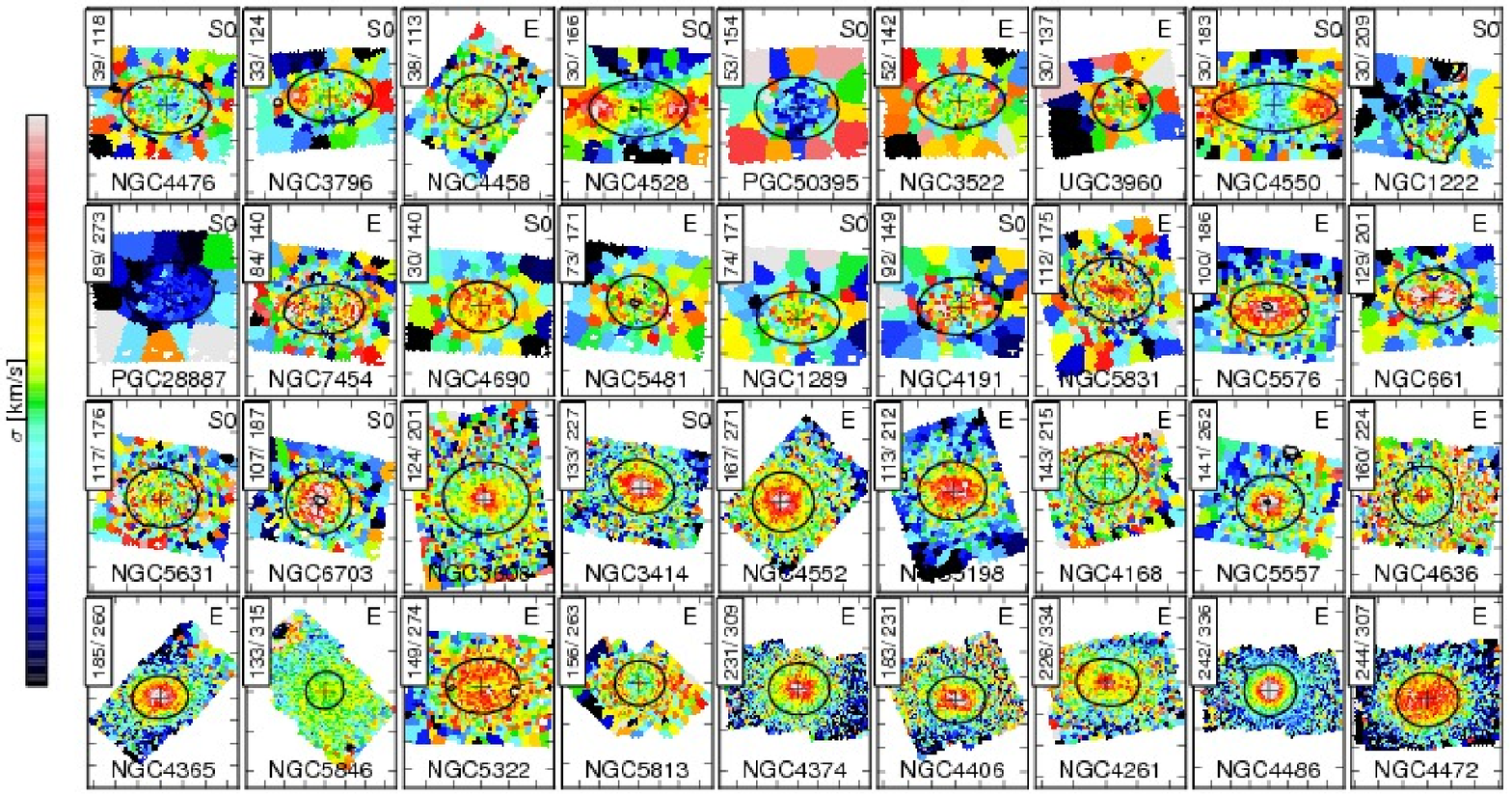, width=\hsize}
\caption{Stellar velocity dispersion fields for the 36 early-type slow rotators in the
\atlas\ sample. See Fig.~\ref{fig:SlowVmaps} for details. The specific colour cuts
adapted to each dispersion map are specified in each inset (in km.s$^{-1}$).}
\label{fig:SlowSmaps}
\end{figure*}

We provide here two figures (Figs.~\ref{fig:SlowVmaps} and \ref{fig:SlowVmaps}) illustrating the stellar velocity and dispersion maps for the 36 slow
rotators of our \atlas\ sample. Galaxies are sorted by (increasing) mass from
left to right, top to bottom.

Fig.~\ref{fig:SlowVmaps} emphasises the non-regularity of
the velocity fields of Slow Rotators, as well as the presence of large-scale
(kpc-size) KDCs. It also illustrates that the galaxies with zero apparent
rotation (NGC\,4374, NGC\,4486 and NGC\,5846) are among the most massive Slow
Rotators. The dispersion maps provided in Fig.~\ref{fig:SlowSmaps} illustrate e.g., how the
stellar velocity dispersion peaks along the major-axis away from the centre for
the Slow Rotators which were labelled "double $\sigma$" in \cite{Krajnovic+11}:
these systems (e.g., NGC\,4550 and NGC\,4528) are on the low-mass end of our
sample.

\section{Dynamical models, $V/\sigma$ and $\lambda_R$}

In E+07 (see their Appendix B), it has been shown that a tight relation exists between $\lambda_R$
and $V/\sigma$ for simple two-integral Jeans models, namely:
\begin{equation}
\lambda_R = \frac{\langle R V \rangle}{\langle R  \sqrt{ V^2 + \sigma^2 }\rangle}
\approx \frac{\kappa \, \left( V / \sigma \right)}{\sqrt{1 + \kappa^2 \, \left( V / \sigma \right)^2}}
\, .
\label{eq:kappa}
\end{equation}
The value of $\kappa$ was estimated both from observations and models to be $\sim 1.1$. 

Here we wish to extend this work both for our complete \atlas\ sample and additional models. 
We have now derived a series of dynamical models for ellipsoidal or spheroidal systems with various intrinsic 
flattening (with ellipticities from 0 to 0.9), luminosity (or mass) profiles (Sersic profiles 
with $n = 2$ or 4), and anisotropy parameters ($\beta$ from 0 to 0.6). 
The dynamical models have an axisymmetric mass distribution described by the MGE parametrization 
\citep{Emsellem+94}, which was derived by fitting a one-dimensional Sersic profile with the MGE 
routines of \cite{Cappellari02}. The models assume a spatially-constant, cylindrically-aligned 
and oblate ($\sigma_{\phi} = \sigma_R$) velocity ellipsoid (where
$\sigma_{\phi}$ and $\sigma_R$ are the azimuthal and radial velocity dispersion,
respectively) characterized by the anisotropy parameter $\beta = 1 - {\left(\sigma_z /
\sigma_R \right)}^2$. Under these assumptions the unique prediction for the two-dimensional $V$ and 
$\sigma$ field for the models was computed by solving the anisotropic Jeans equations with the 
routines of \cite{Cappellari08}. The two-dimensional kinematics of the models was projected at 
different inclinations and was integrated within elliptical apertures of effective radii $R_e$ or $R_e / 2$, 
in the same way as for the observed systems.
\begin{figure}
\centering
\epsfig{file=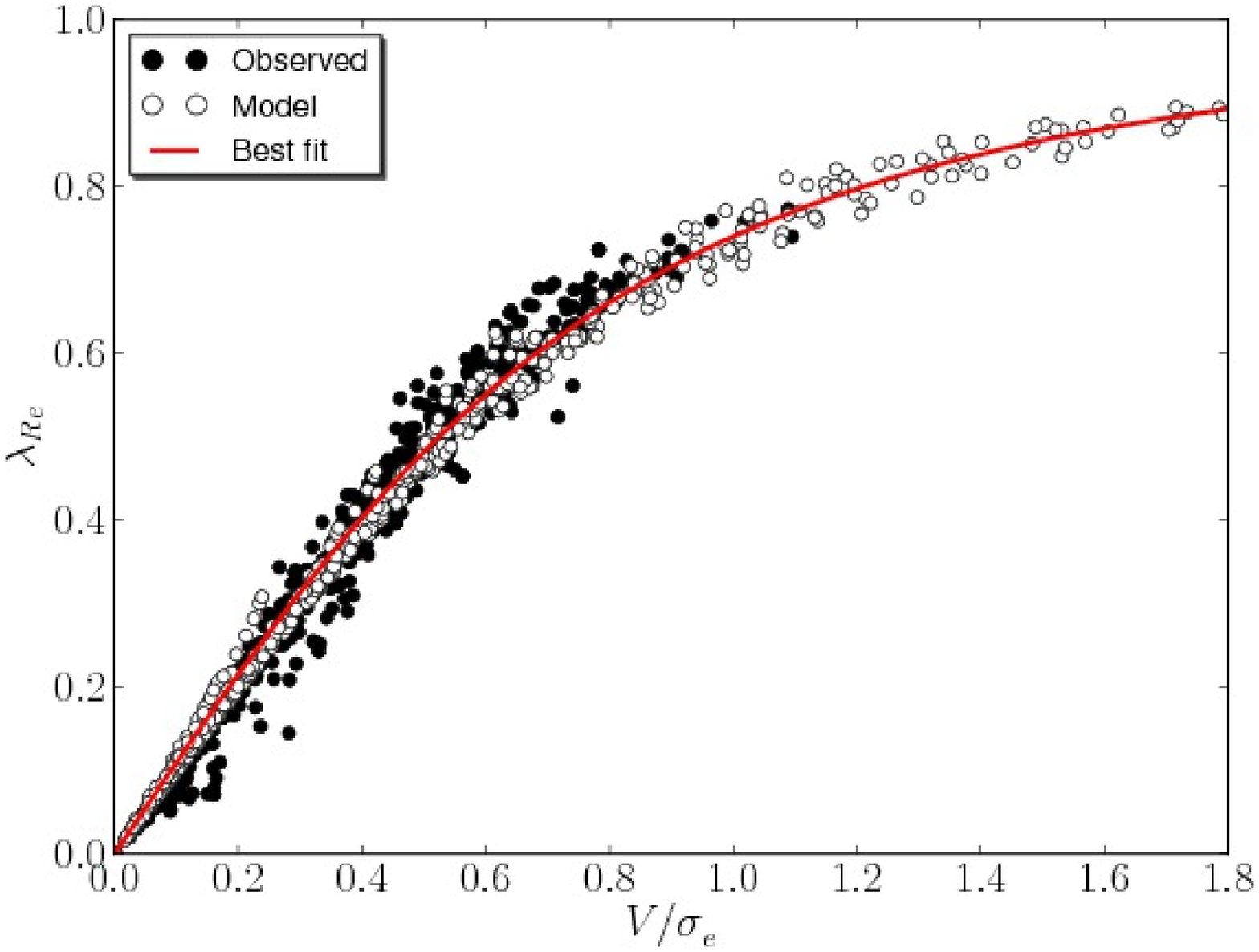, width=\columnwidth}
\caption{$\lambda_R$ versus $V/\sigma$ from the relation given in Eq.\ref{eq:kappa}. 
Results from the dynamical models are shown as open circles, and measurements resulting from the 
\atlas\ observations as black circles. The solid red line is the best fit relation following
Eq.~\ref{eq:kappa} with $\kappa = 1.1$. Note that the observations exhibit a
slightly steeper slope.}
\label{fig:LRVS}
\end{figure}

\begin{figure}
\centering
\epsfig{file=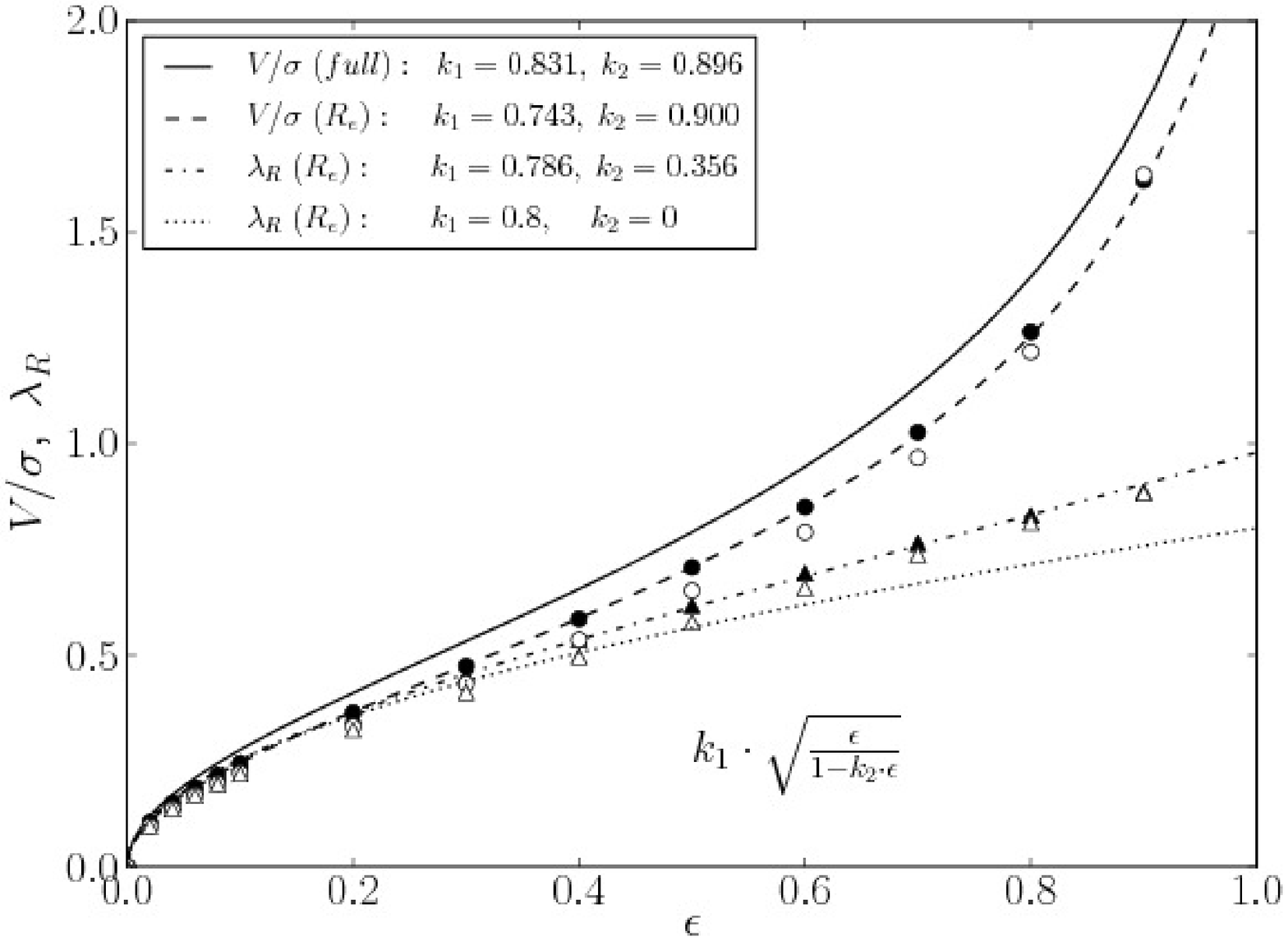, width=\columnwidth}
\caption{$\lambda_R$ (triangles) and $V/\sigma$ (circles) as functions of the ellipticity $\epsilon$ for isotropic systems:
the expected theoretical formula is shown as a solid line \citep[with $\alpha=0.15$, see][for details and notations]{Binney05}, 
while the dashed and
dashed-dotted lines provide the best fits to the edge-on values for $V/\sigma$ and $\lambda_R$ derived within 1~$R_e$, respectively. Filled and open symbols are for mass profiles with Sersic index $n=4$ and $n=2$, respectively. The dotted line shows the approximation using a simple law where $\lambda_R \propto \sqrt{\epsilon}$.}
\label{fig:edgeon}
\end{figure}

\begin{figure}
\centering
\epsfig{file=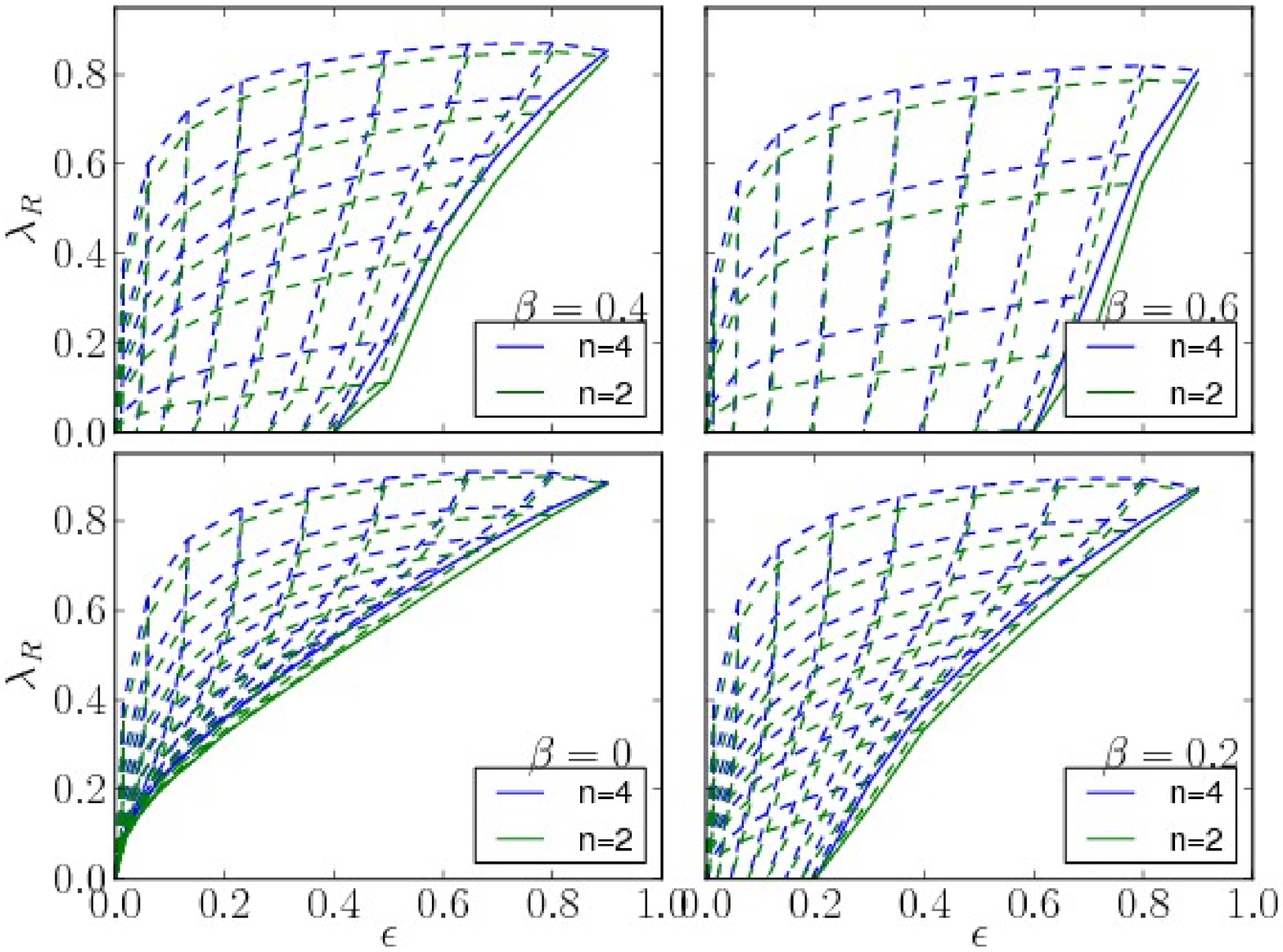, width=\columnwidth}
\caption{$\lambda_R$ versus the ellipticity $\epsilon$ for different values of
the anisotropy parameter $\beta$ and two Sersic profiles ($n=2$ and $n=4$).}
\label{fig:LREPS_aniso}
\end{figure}

\begin{figure}
\centering
\epsfig{file=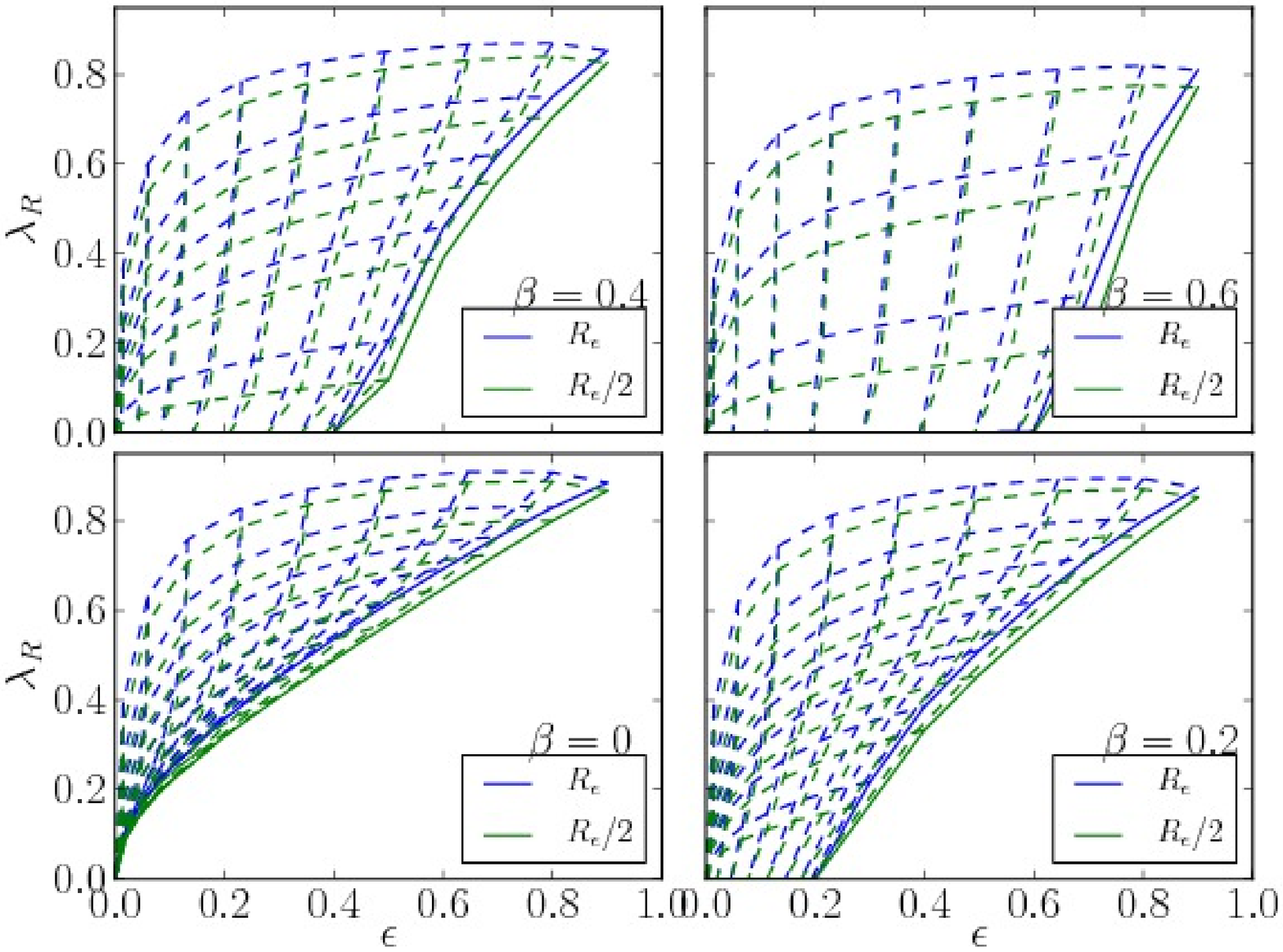, width=\columnwidth}
\caption{$\lambda_R$ versus the ellipticity $\epsilon$ for apertures of 1~$R_e$
and $R_e / 2$.}
\label{fig:LREPS_Re}
\end{figure}
\begin{figure}
\centering
\epsfig{file=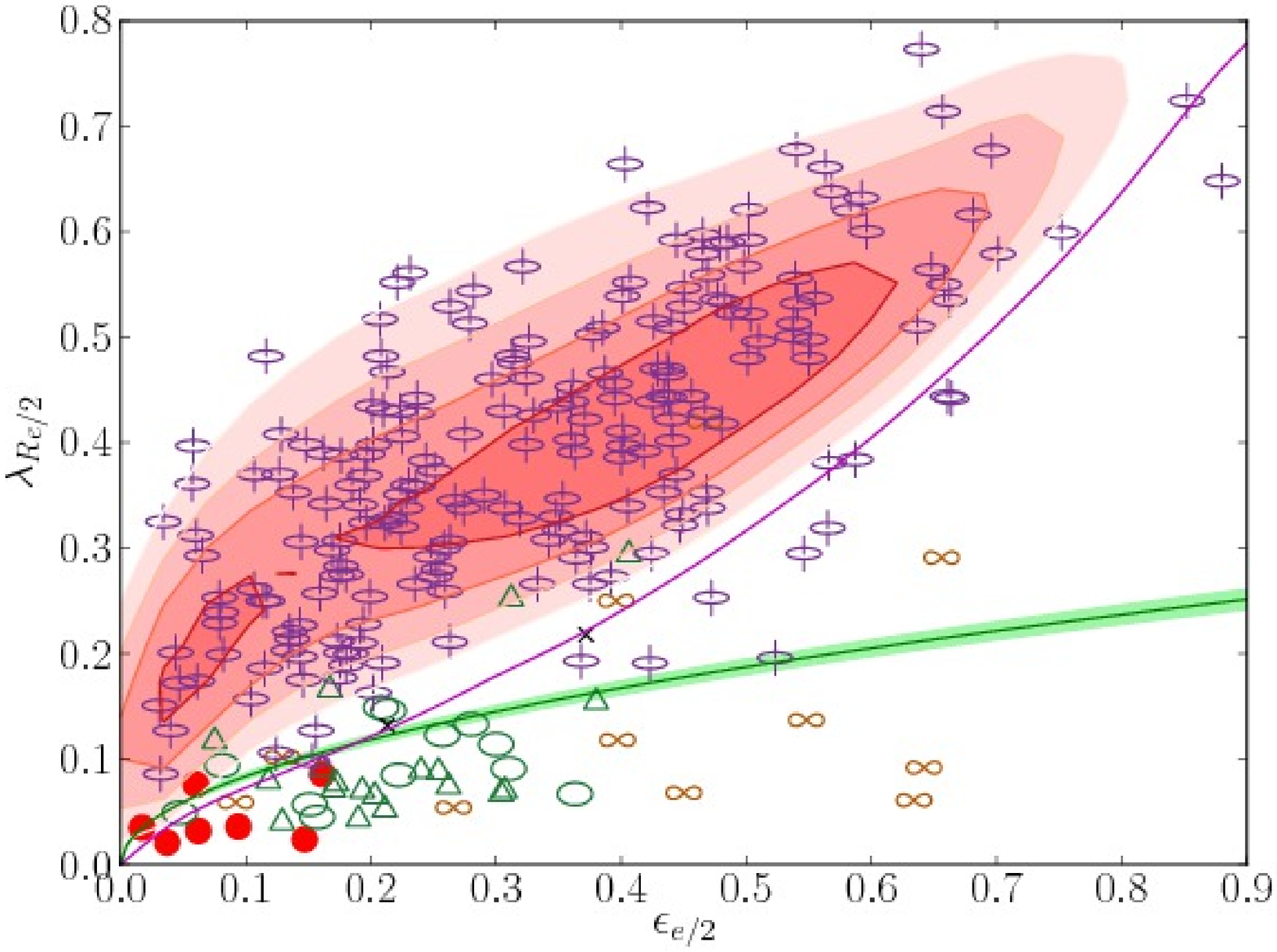, width=\columnwidth}
\caption{${\lambda_R}_{e/2}$ versus apparent ellipticity $\epsilon_{e/2}$ with flags
indicating the observed kinematic structure with symbols as in Fig.~\ref{fig:LRLR2_Flag}. The coloured
contours show the result of a Monte Carlo realisation as described in the text.
The green line shows the limit between Slow and Fast Rotators.}
\label{fig:LR2Eps2_MC}
\end{figure}
We first confirm that $\kappa = 1.1$ for Eq.~\ref{eq:kappa} provides a good fit for both
the models and our complete \atlas\ sample (see Fig.~\ref{fig:LRVS}), although
the slope is slightly steeper for the observed galaxies than for the models, and
the optimal value does in fact vary significantly depending on inclination and
anisotropy. We present the predictions of these models for $\lambda_R$ versus $\epsilon$ for
anisotropy $\beta$ values from 0 to 0.6 in Fig.~\ref{fig:LREPS_aniso}, and for
apertures of 1~$R_e$ and $R_e / 2$. in Fig.~\ref{fig:LREPS_Re}.
Assuming a relation of the form given in Eq.~\ref{eq:kappa}
hold between $V/\sigma$ and $\lambda_R$, we can translate any relation such as 
$\left( V/\sigma \right)(x_1) = C \cdot \left( V/\sigma \right)(x_2)$
by $f_{\lambda_R}(x_1) = C \cdot f_{\lambda_R}(x_2)$ where $f_{\lambda_R} = \lambda_R /
\sqrt{ 1 - {\lambda_R}^2 }$.
We find that generally ${\left(V/\sigma\right)}_e \sim 1.1 \,\, {\left(V/\sigma\right)}_{e/2}$, and
accordingly that $f_{\lambda_R}(R_e) \sim 1.15 \cdot f_{\lambda_R}(R_e/2)$: the
constant of proportionality slightly changes from 1.1 to 1.15 here mainly because
$\kappa$ in Eq.~\ref{eq:kappa} is not exactly constant as we go to smaller apertures.

We then provide the best fit relations for $\lambda_R$ and $V/\sigma$ as
functions of ellipticity $\epsilon$ for isotropic systems, as given
by $k_1 \sqrt{\frac{\epsilon}{1 - k_2 \epsilon}}$.
The theoretical values for $k_1$ and $k_2$ are 0.831 and 0.896, using the entire
dynamical system and assuming $\alpha = 0.15$ \citep[see][]{Binney05}. 
This corresponds to values of $k_1 \sim 0.90$ and $k_2 \sim 0.08$ for
$\lambda_R$. This means that $\lambda_R$ is nearly proportional to
$\sqrt{\epsilon}$ for isotropic systems (with $\alpha=0.15$), leading to a reasonable approximation
for low values of the ellipticity with $\lambda_R \sim 0.91 \times \sqrt{\epsilon}$, and a best fit approximation
over a range $[0-0.9]$ of ellipticities given by $0.93 \, \sqrt{\epsilon}$. 
As shown in C+07, the expected values using an effective
aperture of $R_e / 2$ are slightly smaller but would be close to the theoretical
values obtained with $\alpha \sim 0.2$. The best fit relation for edge-on values
within 1~$R_e$ gives $k_1 = 0.743$ and $k_2 = 0.900$ for $V/\sigma$. There is a
slight dependency of these parameters when the mass profile is changed (Sersic
index $n=4$ and $n=2$), as expected. 

The corresponding (edge-on) best fit relation for $\lambda_R$ provides $k_1 = 0.786$ and $k_2 = 0.355$
for apertures of $1~R_e$, and $k_1 = 0.69$ and $k_2 = 0.50$ for $R_e / 2$.
Restricting ourselves to the lower range of ellipticities (below 0.35), we have
$\lambda_R \sim 0.8 \times \sqrt{\epsilon}$ and $\lambda_R \sim 0.71 \times \sqrt{\epsilon}$ for apertures
of 1~$R_e$ and $R_e/2$, respectively. 

It is worth mentioning the effect on inclination for both $V/\sigma$
and $\lambda_R$ values. As shown by \cite{Binney05}, for an inclination $i$, we
have ($i=90\degr$ being edge-on):
$$
\left({V/\sigma}\right)(i) = C(i) \cdot \left({V/\sigma}\right)_{edge-on}
$$
with $C(i) = \sin{i} / \sqrt{1 - \beta\,\cos^2{i}}$ where we assumed an
axisymmetric system for which the intersection between the velocity ellipsoid and a
plane perpendicular to the axis of symmetry is a circle everywhere. Assuming the
relation \ref{eq:kappa} holds, this translates into:
$$
{\lambda_R}(i) = C(i) \cdot \frac{{\lambda_R}_{edge-on} }{\sqrt{1 + (C^2(i)-1)\cdot {\lambda_R}^2_{edge-on}}}
$$

We finally include in Fig.~\ref{fig:LR2Eps2_MC} a Monte Carlo simulation as in Fig.~\ref{fig:LREps_MC}
(see Sect.~\ref{sec:MC}) but here for a smaller aperture of $R_e/2$ for comparison.
We use a distribution for the intrinsic ellipticities $\epsilon_{intr}$ of the simulated sample
with two Gaussians: one centred at $\epsilon_{0} = 0.7$ with a width $\sigma_{\epsilon} = 0.2$ as
for the simulation at 1~$R_e$, and we add one component centred at $\epsilon_0 = 0.6$ 
representing 30\% of the one at 0.7. This second Gaussian is required to account
for the fact that for an aperture of $R_e / 2$ the intrinsic ellipticity distribution
is offset to values sligthly smaller than for an aperture size of $R_e$, the larger one
better probing the outer more flattened component of (most) Fast Rotators.
We again fix the anisotropy distribution also as a Gaussian with a mean of 
$m_{\beta} = 0.5$ and a dispersion of $\sigma_{\beta}
= 0.1$, truncated at $0.8 \epsilon_{intr}$.

\newpage
\clearpage

\begin{deluxetable}{lcccccccccc} 
  \tablewidth{0pt} 
  \setlength{\tabcolsep}{2pt} 
  \tablecaption{Ellipticities, $\lambda_R$ and $V/\sigma$ tabulated values for the \atlas\ sample or 260 early-type galaxies} 
  \tablehead{ 
  \colhead{Name} & 
  \colhead{$R_{max}$} & 
  \colhead{$\epsilon_e$} & 
  \colhead{$\epsilon_{e/2}$} & 
  \colhead{$Band({\epsilon})$} & 
  \colhead{${V/\sigma}_{e}$} & 
  \colhead{${V/\sigma}_{e/2}$} & 
  \colhead{${\lambda_R}_{e}$} & 
  \colhead{${\lambda_R}_{e/2}$} & 
  \colhead{F/S ($R_e$)} & 
  \colhead{F/S ($R_e / 2$)} \\ 
  \colhead{} & 
  \colhead{} & 
  \colhead{} & 
  \colhead{} & 
  \colhead{} & 
  \colhead{} & 
  \colhead{} & 
  \colhead{} & 
  \colhead{} & 
  \colhead{} & 
  \colhead{} \\ 
  \colhead{(1)} & 
  \colhead{(2)} & 
  \colhead{(3)} & 
  \colhead{(4)} & 
  \colhead{(5)} & 
  \colhead{(6)} & 
  \colhead{(7)} & 
  \colhead{(8)} & 
  \colhead{(9)} & 
  \colhead{(10)} & 
  \colhead{(11)} 
  } 
  \startdata
 IC0560 &     1.10 &    0.587 &    0.479 & G &     0.770 &     0.607 &      0.691 &     0.589 &  F & F\\
 IC0598 &     1.39 &    0.579 &    0.502 & G &     0.896 &     0.629 &      0.736 &     0.592 &  F & F\\
 IC0676 &     0.68 &    0.524 &    0.664 & G &     0.508 &     0.436 &      0.495 &     0.442 &  F & F\\
 IC0719 &     1.06 &    0.714 &    0.642 & R &     0.279 &     0.117 &      0.263 &     0.092 &  F & S\\
 IC0782 &     0.76 &    0.404 &    0.263 & G &     0.677 &     0.608 &      0.571 &     0.529 &  F & F\\
 IC1024 &     1.18 &    0.679 &    0.752 & G &     0.916 &     0.647 &      0.724 &     0.599 &  F & F\\
 IC3631 &     1.11 &    0.560 &    0.524 & G &     0.218 &     0.173 &      0.250 &     0.196 &  F & F\\
NGC0448 &     1.22 &    0.643 &    0.656 & G &     0.380 &     0.377 &      0.327 &     0.291 &  F & F\\
NGC0474 &     0.52 &    0.192 &    0.191 & G &     0.229 &     0.227 &      0.213 &     0.210 &  F & F\\
NGC0502 &     1.44 &    0.146 &    0.047 & G &     0.213 &     0.168 &      0.234 &     0.172 &  F & F\\
NGC0509 &     0.56 &    0.679 &    0.661 & G &     0.451 &     0.430 &      0.465 &     0.444 &  F & F\\
NGC0516 &     0.95 &    0.699 &    0.701 & G &     0.786 &     0.624 &      0.655 &     0.579 &  F & F\\
NGC0524 &     0.50 &    0.034 &    0.034 & G &     0.344 &     0.344 &      0.325 &     0.325 &  F & F\\
NGC0525 &     1.41 &    0.280 &    0.173 & G &     0.404 &     0.285 &      0.418 &     0.274 &  F & F\\
NGC0661 &     1.33 &    0.306 &    0.307 & G &     0.134 &     0.094 &      0.139 &     0.070 &  S & S\\
NGC0680 &     1.19 &    0.189 &    0.185 & G &     0.413 &     0.361 &      0.406 &     0.360 &  F & F\\
NGC0770 &     2.01 &    0.276 &    0.299 & G &     0.184 &     0.117 &      0.195 &     0.114 &  F & S\\
NGC0821 &     0.49 &    0.392 &    0.392 & G &     0.288 &     0.288 &      0.273 &     0.273 &  F & F\\
NGC0936 &     1.00 &    0.223 &    0.223 & G &     0.677 &     0.448 &      0.657 &     0.430 &  F & F\\
NGC1023 &     0.47 &    0.363 &    0.363 & G &     0.372 &     0.372 &      0.391 &     0.391 &  F & F\\
NGC1121 &     2.14 &    0.559 &    0.450 & G &     0.828 &     0.587 &      0.711 &     0.548 &  F & F\\
NGC1222 &     1.12 &    0.280 &    0.280 & G &     0.153 &     0.138 &      0.147 &     0.132 &  S & S\\
NGC1248 &     1.10 &    0.227 &    0.138 & G &     0.521 &     0.337 &      0.535 &     0.353 &  F & F\\
NGC1266 &     0.88 &    0.193 &    0.207 & G &     0.775 &     0.589 &      0.638 &     0.517 &  F & F\\
NGC1289 &     0.90 &    0.393 &    0.380 & G &     0.172 &     0.154 &      0.178 &     0.154 &  S & S\\
NGC1665 &     0.65 &    0.491 &    0.440 & G &     0.669 &     0.478 &      0.658 &     0.511 &  F & F\\
NGC2481 &     1.26 &    0.684 &    0.467 & G &     0.686 &     0.399 &      0.678 &     0.428 &  F & F\\
NGC2549 &     0.88 &    0.587 &    0.488 & G &     0.623 &     0.563 &      0.572 &     0.523 &  F & F\\
NGC2577 &     1.13 &    0.471 &    0.421 & G &     0.874 &     0.733 &      0.696 &     0.623 &  F & F\\
NGC2592 &     1.39 &    0.208 &    0.210 & G &     0.529 &     0.422 &      0.549 &     0.431 &  F & F\\
NGC2594 &     2.92 &    0.403 &    0.433 & G &     0.455 &     0.450 &      0.404 &     0.444 &  F & F\\
NGC2679 &     0.46 &    0.368 &    0.368 & G &     0.185 &     0.185 &      0.193 &     0.193 &  F & F\\
NGC2685 &     0.90 &    0.612 &    0.592 & G &     0.840 &     0.734 &      0.691 &     0.632 &  F & F\\
NGC2695 &     1.04 &    0.293 &    0.225 & G &     0.582 &     0.422 &      0.575 &     0.406 &  F & F\\
NGC2698 &     1.27 &    0.516 &    0.267 & S &     0.569 &     0.331 &      0.593 &     0.345 &  F & F\\
NGC2699 &     1.39 &    0.143 &    0.202 & G &     0.450 &     0.393 &      0.444 &     0.398 &  F & F\\
NGC2764 &     1.25 &    0.614 &    0.662 & G &     0.783 &     0.568 &      0.658 &     0.535 &  F & F\\
NGC2768 &     0.35 &    0.472 &    0.472 & G &     0.236 &     0.236 &      0.253 &     0.253 &  F & F\\
NGC2778 &     1.05 &    0.224 &    0.201 & G &     0.574 &     0.419 &      0.572 &     0.435 &  F & F\\
NGC2824 &     2.49 &    0.162 &    0.107 & G &     0.534 &     0.407 &      0.492 &     0.370 &  F & F\\
NGC2852 &     2.60 &    0.135 &    0.043 & G &     0.320 &     0.186 &      0.368 &     0.201 &  F & F\\
NGC2859 &     0.60 &    0.089 &    0.057 & G &     0.410 &     0.408 &      0.359 &     0.361 &  F & F\\
NGC2880 &     0.81 &    0.282 &    0.208 & G &     0.627 &     0.508 &      0.581 &     0.482 &  F & F\\
NGC2950 &     1.11 &    0.362 &    0.237 & G &     0.488 &     0.472 &      0.436 &     0.428 &  F & F\\
NGC2962 &     0.64 &    0.484 &    0.319 & G &     0.376 &     0.298 &      0.430 &     0.329 &  F & F\\
NGC2974 &     0.53 &    0.407 &    0.403 & R &     0.813 &     0.811 &      0.663 &     0.664 &  F & F\\
NGC3032 &     1.32 &    0.102 &    0.176 & G &     0.267 &     0.173 &      0.344 &     0.201 &  F & F\\
NGC3073 &     1.24 &    0.124 &    0.162 & G &     0.136 &     0.092 &      0.145 &     0.086 &  F & S\\
NGC3098 &     1.20 &    0.553 &    0.440 & S &     0.744 &     0.446 &      0.635 &     0.422 &  F & F\\
NGC3156 &     0.96 &    0.478 &    0.469 & G &     0.771 &     0.611 &      0.648 &     0.559 &  F & F\\
NGC3182 &     0.79 &    0.166 &    0.195 & G &     0.372 &     0.405 &      0.306 &     0.369 &  F & F\\
NGC3193 &     0.57 &    0.129 &    0.143 & G &     0.208 &     0.208 &      0.195 &     0.197 &  F & F\\
NGC3226 &     0.53 &    0.168 &    0.159 & G &     0.273 &     0.275 &      0.251 &     0.257 &  F & F\\
NGC3230 &     0.81 &    0.550 &    0.360 & G &     0.617 &     0.393 &      0.609 &     0.403 &  F & F\\
NGC3245 &     0.62 &    0.442 &    0.444 & G &     0.624 &     0.579 &      0.626 &     0.592 &  F & F\\
NGC3248 &     1.09 &    0.374 &    0.290 & G &     0.514 &     0.380 &      0.511 &     0.350 &  F & F\\
NGC3301 &     0.81 &    0.507 &    0.324 & G &     0.676 &     0.550 &      0.603 &     0.461 &  F & F\\
NGC3377 &     0.42 &    0.503 &    0.503 & G &     0.564 &     0.564 &      0.522 &     0.522 &  F & F\\
NGC3379 &     0.50 &    0.104 &    0.104 & G &     0.152 &     0.152 &      0.157 &     0.157 &  F & F\\
NGC3384 &     0.54 &    0.065 &    0.058 & G &     0.456 &     0.449 &      0.407 &     0.397 &  F & F\\
NGC3400 &     1.02 &    0.437 &    0.437 & G &     0.764 &     0.480 &      0.664 &     0.470 &  F & F\\
NGC3412 &     0.51 &    0.441 &    0.441 & G &     0.375 &     0.371 &      0.406 &     0.403 &  F & F\\
NGC3414 &     0.67 &    0.194 &    0.193 & G &     0.106 &     0.107 &      0.073 &     0.070 &  S & S\\
NGC3457 &     1.28 &    0.033 &    0.032 & S &     0.109 &     0.091 &      0.104 &     0.086 &  F & F\\
NGC3458 &     1.61 &    0.231 &    0.114 & G &     0.438 &     0.245 &      0.461 &     0.250 &  F & F\\
NGC3489 &     0.77 &    0.262 &    0.221 & G &     0.687 &     0.621 &      0.589 &     0.552 &  F & F\\
NGC3499 &     2.03 &    0.139 &    0.143 & G &     0.323 &     0.236 &      0.318 &     0.227 &  F & F\\
NGC3522 &     1.75 &    0.361 &    0.262 & S &     0.080 &     0.091 &      0.058 &     0.074 &  S & S\\
NGC3530 &     1.91 &    0.642 &    0.555 & G &     0.741 &     0.554 &      0.656 &     0.537 &  F & F\\
NGC3595 &     1.10 &    0.507 &    0.376 & G &     0.456 &     0.278 &      0.510 &     0.301 &  F & F\\
NGC3599 &     0.65 &    0.080 &    0.080 & G &     0.273 &     0.241 &      0.282 &     0.239 &  F & F\\
NGC3605 &     0.95 &    0.409 &    0.353 & R &     0.415 &     0.336 &      0.434 &     0.347 &  F & F\\
NGC3607 &     0.60 &    0.185 &    0.194 & G &     0.283 &     0.294 &      0.209 &     0.228 &  F & F\\
NGC3608 &     0.49 &    0.190 &    0.190 & G &     0.054 &     0.054 &      0.043 &     0.043 &  S & S\\
NGC3610 &     1.10 &    0.381 &    0.400 & G &     0.642 &     0.639 &      0.530 &     0.539 &  F & F\\
NGC3613 &     0.59 &    0.460 &    0.423 & G &     0.202 &     0.176 &      0.222 &     0.191 &  F & F\\
NGC3619 &     0.62 &    0.173 &    0.176 & G &     0.285 &     0.274 &      0.293 &     0.283 &  F & F\\
NGC3626 &     0.57 &    0.591 &    0.583 & G &     0.728 &     0.681 &      0.660 &     0.620 &  F & F\\
NGC3630 &     1.04 &    0.665 &    0.371 & G &     0.639 &     0.404 &      0.648 &     0.421 &  F & F\\
NGC3640 &     0.76 &    0.191 &    0.224 & G &     0.370 &     0.315 &      0.377 &     0.320 &  F & F\\
NGC3641 &     1.84 &    0.125 &    0.215 & G &     0.331 &     0.353 &      0.243 &     0.328 &  F & F\\
NGC3648 &     1.24 &    0.421 &    0.326 & G &     0.711 &     0.479 &      0.684 &     0.496 &  F & F\\
NGC3658 &     0.91 &    0.243 &    0.148 & G &     0.462 &     0.336 &      0.546 &     0.398 &  F & F\\
NGC3665 &     0.77 &    0.216 &    0.176 & G &     0.463 &     0.439 &      0.410 &     0.388 &  F & F\\
NGC3674 &     1.28 &    0.568 &    0.342 & G &     0.491 &     0.292 &      0.541 &     0.308 &  F & F\\
NGC3694 &     1.71 &    0.268 &    0.208 & G &     0.269 &     0.203 &      0.256 &     0.191 &  F & F\\
NGC3757 &     2.06 &    0.153 &    0.153 & G &     0.119 &     0.096 &      0.126 &     0.098 &  F & S\\
NGC3796 &     1.53 &    0.361 &    0.398 & G &     0.164 &     0.111 &      0.171 &     0.119 &  S & S\\
NGC3838 &     1.42 &    0.605 &    0.385 & S &     0.743 &     0.536 &      0.676 &     0.509 &  F & F\\
NGC3941 &     0.66 &    0.251 &    0.251 & G &     0.439 &     0.352 &      0.468 &     0.373 &  F & F\\
NGC3945 &     0.74 &    0.090 &    0.230 & G &     0.717 &     0.741 &      0.524 &     0.561 &  F & F\\
NGC3998 &     0.87 &    0.170 &    0.164 & G &     0.415 &     0.326 &      0.454 &     0.342 &  F & F\\
NGC4026 &     0.75 &    0.556 &    0.438 & G &     0.540 &     0.451 &      0.548 &     0.442 &  F & F\\
NGC4036 &     0.53 &    0.555 &    0.540 & G &     0.818 &     0.803 &      0.685 &     0.678 &  F & F\\
NGC4078 &     2.00 &    0.561 &    0.510 & S &     0.575 &     0.524 &      0.523 &     0.496 &  F & F\\
NGC4111 &     1.25 &    0.582 &    0.498 & S &     0.720 &     0.663 &      0.619 &     0.567 &  F & F\\
NGC4119 &     0.61 &    0.598 &    0.563 & G &     0.895 &     0.808 &      0.701 &     0.660 &  F & F\\
NGC4143 &     0.73 &    0.390 &    0.324 & G &     0.504 &     0.369 &      0.538 &     0.398 &  F & F\\
NGC4150 &     0.98 &    0.328 &    0.274 & G &     0.514 &     0.337 &      0.513 &     0.338 &  F & F\\
NGC4168 &     0.46 &    0.129 &    0.129 & G &     0.047 &     0.047 &      0.040 &     0.040 &  S & S\\
NGC4179 &     0.77 &    0.591 &    0.501 & G &     0.589 &     0.470 &      0.587 &     0.480 &  F & F\\
NGC4191 &     1.54 &    0.269 &    0.092 & G &     0.105 &     0.073 &      0.107 &     0.059 &  S & S\\
NGC4203 &     0.59 &    0.154 &    0.180 & G &     0.281 &     0.255 &      0.305 &     0.275 &  F & F\\
NGC4215 &     0.83 &    0.702 &    0.444 & G &     0.516 &     0.363 &      0.553 &     0.370 &  F & F\\
NGC4233 &     1.15 &    0.277 &    0.162 & G &     0.578 &     0.401 &      0.564 &     0.390 &  F & F\\
NGC4249 &     1.32 &    0.048 &    0.040 & G &     0.204 &     0.142 &      0.194 &     0.127 &  F & F\\
NGC4251 &     0.79 &    0.508 &    0.387 & G &     0.637 &     0.503 &      0.584 &     0.466 &  F & F\\
NGC4255 &     1.34 &    0.434 &    0.127 & S &     0.622 &     0.374 &      0.606 &     0.370 &  F & F\\
NGC4259 &     1.93 &    0.553 &    0.450 & G &     0.231 &     0.081 &      0.237 &     0.068 &  F & S\\
NGC4261 &     0.44 &    0.222 &    0.222 & G &     0.090 &     0.090 &      0.085 &     0.085 &  S & S\\
NGC4262 &     1.39 &    0.117 &    0.117 & G &     0.297 &     0.254 &      0.309 &     0.250 &  F & F\\
NGC4264 &     1.26 &    0.191 &    0.191 & G &     0.541 &     0.355 &      0.499 &     0.341 &  F & F\\
NGC4267 &     0.53 &    0.079 &    0.079 & G &     0.260 &     0.241 &      0.282 &     0.253 &  F & F\\
NGC4268 &     0.93 &    0.552 &    0.348 & G &     0.491 &     0.333 &      0.492 &     0.330 &  F & F\\
NGC4270 &     0.96 &    0.543 &    0.424 & G &     0.378 &     0.289 &      0.406 &     0.294 &  F & F\\
NGC4278 &     0.72 &    0.103 &    0.134 & G &     0.201 &     0.208 &      0.178 &     0.203 &  F & F\\
NGC4281 &     0.69 &    0.535 &    0.502 & G &     0.757 &     0.702 &      0.651 &     0.621 &  F & F\\
NGC4283 &     1.44 &    0.033 &    0.031 & G &     0.172 &     0.155 &      0.171 &     0.151 &  F & F\\
NGC4324 &     0.83 &    0.434 &    0.197 & G &     0.639 &     0.407 &      0.592 &     0.389 &  F & F\\
NGC4339 &     0.55 &    0.060 &    0.060 & G &     0.311 &     0.302 &      0.325 &     0.312 &  F & F\\
NGC4340 &     0.62 &    0.227 &    0.237 & G &     0.499 &     0.481 &      0.458 &     0.442 &  F & F\\
NGC4342 &     2.45 &    0.442 &    0.263 & S &     0.537 &     0.307 &      0.528 &     0.306 &  F & F\\
NGC4346 &     0.77 &    0.533 &    0.360 & G &     0.571 &     0.428 &      0.576 &     0.439 &  F & F\\
NGC4350 &     0.77 &    0.674 &    0.550 & G &     0.658 &     0.460 &      0.638 &     0.480 &  F & F\\
NGC4365 &     0.33 &    0.254 &    0.254 & G &     0.114 &     0.114 &      0.088 &     0.088 &  S & S\\
NGC4371 &     0.51 &    0.309 &    0.313 & G &     0.541 &     0.541 &      0.482 &     0.482 &  F & F\\
NGC4374 &     0.44 &    0.147 &    0.147 & G &     0.026 &     0.026 &      0.024 &     0.024 &  S & S\\
NGC4377 &     1.30 &    0.157 &    0.228 & S &     0.443 &     0.314 &      0.473 &     0.338 &  F & F\\
NGC4379 &     0.90 &    0.224 &    0.258 & G &     0.386 &     0.294 &      0.398 &     0.300 &  F & F\\
NGC4382 &     0.34 &    0.202 &    0.202 & G &     0.174 &     0.174 &      0.163 &     0.163 &  F & F\\
NGC4387 &     1.10 &    0.373 &    0.351 & G &     0.391 &     0.315 &      0.399 &     0.317 &  F & F\\
NGC4406 &     0.24 &    0.211 &    0.211 & G &     0.090 &     0.090 &      0.052 &     0.052 &  S & S\\
NGC4417 &     0.89 &    0.566 &    0.418 & G &     0.542 &     0.395 &      0.548 &     0.392 &  F & F\\
NGC4425 &     0.62 &    0.676 &    0.587 & G &     0.438 &     0.363 &      0.454 &     0.384 &  F & F\\
NGC4429 &     0.48 &    0.402 &    0.402 & G &     0.455 &     0.455 &      0.396 &     0.396 &  F & F\\
NGC4434 &     1.15 &    0.058 &    0.082 & G &     0.249 &     0.186 &      0.288 &     0.199 &  F & F\\
NGC4435 &     0.52 &    0.468 &    0.465 & G &     0.672 &     0.667 &      0.599 &     0.597 &  F & F\\
NGC4442 &     0.62 &    0.335 &    0.307 & G &     0.350 &     0.329 &      0.363 &     0.338 &  F & F\\
NGC4452 &     0.42 &    0.880 &    0.880 & G &     0.765 &     0.765 &      0.648 &     0.648 &  F & F\\
NGC4458 &     0.68 &    0.121 &    0.119 & G &     0.150 &     0.161 &      0.072 &     0.079 &  S & S\\
NGC4459 &     0.64 &    0.148 &    0.128 & G &     0.477 &     0.450 &      0.438 &     0.408 &  F & F\\
NGC4461 &     0.68 &    0.392 &    0.306 & G &     0.486 &     0.405 &      0.511 &     0.430 &  F & F\\
NGC4472 &     0.26 &    0.172 &    0.172 & G &     0.073 &     0.073 &      0.077 &     0.077 &  S & S\\
NGC4473 &     0.71 &    0.421 &    0.396 & G &     0.255 &     0.263 &      0.229 &     0.250 &  F & F\\
NGC4474 &     0.72 &    0.570 &    0.468 & G &     0.439 &     0.347 &      0.446 &     0.353 &  F & F\\
NGC4476 &     1.07 &    0.353 &    0.375 & G &     0.236 &     0.298 &      0.153 &     0.266 &  S & F\\
NGC4477 &     0.44 &    0.135 &    0.135 & G &     0.216 &     0.216 &      0.221 &     0.221 &  F & F\\
NGC4478 &     0.98 &    0.165 &    0.175 & G &     0.221 &     0.175 &      0.233 &     0.177 &  F & F\\
NGC4483 &     1.02 &    0.346 &    0.251 & G &     0.474 &     0.271 &      0.465 &     0.273 &  F & F\\
NGC4486 &     0.28 &    0.037 &    0.037 & G &     0.024 &     0.024 &      0.021 &     0.021 &  S & S\\
NGC4486A &     1.96 &    0.224 &    0.224 & G &     0.379 &     0.379 &      0.351 &     0.351 &  F & F\\
NGC4489 &     0.62 &    0.085 &    0.075 & G &     0.176 &     0.125 &      0.168 &     0.117 &  F & F\\
NGC4494 &     0.48 &    0.173 &    0.173 & G &     0.219 &     0.219 &      0.212 &     0.212 &  F & F\\
NGC4503 &     0.61 &    0.446 &    0.429 & G &     0.510 &     0.451 &      0.523 &     0.470 &  F & F\\
NGC4521 &     0.89 &    0.566 &    0.482 & G &     0.794 &     0.566 &      0.682 &     0.534 &  F & F\\
NGC4526 &     0.45 &    0.361 &    0.361 & G &     0.563 &     0.563 &      0.453 &     0.453 &  F & F\\
NGC4528 &     1.15 &    0.392 &    0.129 & G &     0.112 &     0.116 &      0.096 &     0.102 &  S & F\\
NGC4546 &     0.66 &    0.527 &    0.465 & G &     0.653 &     0.572 &      0.639 &     0.579 &  F & F\\
NGC4550 &     0.88 &    0.649 &    0.633 & G &     0.122 &     0.072 &      0.105 &     0.061 &  S & S\\
NGC4551 &     0.92 &    0.284 &    0.259 & G &     0.311 &     0.285 &      0.295 &     0.259 &  F & F\\
NGC4552 &     0.46 &    0.047 &    0.047 & G &     0.051 &     0.051 &      0.049 &     0.049 &  S & S\\
NGC4564 &     0.78 &    0.560 &    0.477 & G &     0.640 &     0.532 &      0.619 &     0.536 &  F & F\\
NGC4570 &     0.70 &    0.626 &    0.551 & G &     0.586 &     0.470 &      0.603 &     0.498 &  F & F\\
NGC4578 &     0.70 &    0.289 &    0.282 & G &     0.669 &     0.596 &      0.596 &     0.544 &  F & F\\
NGC4596 &     0.56 &    0.254 &    0.254 & G &     0.310 &     0.297 &      0.297 &     0.280 &  F & F\\
NGC4608 &     0.56 &    0.115 &    0.115 & G &     0.194 &     0.180 &      0.203 &     0.185 &  F & F\\
NGC4612 &     0.75 &    0.204 &    0.196 & R &     0.428 &     0.345 &      0.407 &     0.324 &  F & F\\
NGC4621 &     0.50 &    0.365 &    0.364 & G &     0.273 &     0.272 &      0.291 &     0.291 &  F & F\\
NGC4623 &     0.64 &    0.673 &    0.648 & G &     0.660 &     0.602 &      0.598 &     0.564 &  F & F\\
NGC4624 &     0.42 &    0.065 &    0.065 & G &     0.292 &     0.288 &      0.297 &     0.293 &  F & F\\
NGC4636 &     0.25 &    0.094 &    0.094 & G &     0.039 &     0.039 &      0.036 &     0.036 &  S & S\\
NGC4638 &     0.87 &    0.606 &    0.657 & G &     0.909 &     0.895 &      0.691 &     0.715 &  F & F\\
NGC4643 &     0.61 &    0.249 &    0.199 & G &     0.321 &     0.323 &      0.255 &     0.254 &  F & F\\
NGC4649 &     0.35 &    0.156 &    0.156 & G &     0.123 &     0.123 &      0.127 &     0.127 &  F & F\\
NGC4660 &     1.39 &    0.441 &    0.315 & G &     0.603 &     0.518 &      0.553 &     0.475 &  F & F\\
NGC4684 &     0.70 &    0.598 &    0.596 & G &     0.713 &     0.665 &      0.622 &     0.600 &  F & F\\
NGC4690 &     0.96 &    0.266 &    0.257 & G &     0.149 &     0.130 &      0.151 &     0.123 &  S & S\\
NGC4694 &     0.52 &    0.547 &    0.546 & G &     0.275 &     0.276 &      0.289 &     0.295 &  F & F\\
NGC4697 &     0.37 &    0.447 &    0.447 & G &     0.363 &     0.363 &      0.322 &     0.322 &  F & F\\
NGC4710 &     0.74 &    0.699 &    0.395 & G &     0.731 &     0.496 &      0.652 &     0.456 &  F & F\\
NGC4733 &     0.74 &    0.060 &    0.060 & G &     0.096 &     0.088 &      0.086 &     0.076 &  F & F\\
NGC4753 &     0.49 &    0.213 &    0.213 & G &     0.537 &     0.537 &      0.467 &     0.467 &  F & F\\
NGC4754 &     0.68 &    0.480 &    0.480 & G &     0.445 &     0.395 &      0.467 &     0.418 &  F & F\\
NGC4762 &     0.40 &    0.852 &    0.852 & G &     0.783 &     0.783 &      0.724 &     0.724 &  F & F\\
NGC4803 &     2.02 &    0.282 &    0.266 & S &     0.192 &     0.063 &      0.181 &     0.054 &  F & S\\
NGC5103 &     1.47 &    0.594 &    0.400 & G &     0.588 &     0.370 &      0.573 &     0.386 &  F & F\\
NGC5173 &     1.71 &    0.133 &    0.124 & G &     0.164 &     0.110 &      0.168 &     0.106 &  F & F\\
NGC5198 &     0.62 &    0.146 &    0.151 & G &     0.074 &     0.072 &      0.061 &     0.057 &  S & S\\
NGC5273 &     0.61 &    0.108 &    0.116 & G &     0.554 &     0.512 &      0.517 &     0.482 &  F & F\\
NGC5308 &     0.78 &    0.735 &    0.637 & G &     0.641 &     0.478 &      0.651 &     0.510 &  F & F\\
NGC5322 &     0.56 &    0.307 &    0.304 & G &     0.126 &     0.121 &      0.073 &     0.067 &  S & S\\
NGC5342 &     1.51 &    0.586 &    0.425 & G &     0.637 &     0.410 &      0.626 &     0.439 &  F & F\\
NGC5353 &     0.77 &    0.552 &    0.541 & G &     0.689 &     0.544 &      0.618 &     0.532 &  F & F\\
NGC5355 &     1.59 &    0.296 &    0.263 & G &     0.288 &     0.214 &      0.286 &     0.211 &  F & F\\
NGC5358 &     1.32 &    0.592 &    0.395 & G &     0.675 &     0.461 &      0.610 &     0.442 &  F & F\\
NGC5379 &     0.67 &    0.627 &    0.696 & G &     0.920 &     0.794 &      0.719 &     0.677 &  F & F\\
NGC5422 &     0.68 &    0.604 &    0.537 & G &     0.588 &     0.475 &      0.600 &     0.501 &  F & F\\
NGC5473 &     0.71 &    0.211 &    0.211 & G &     0.437 &     0.326 &      0.447 &     0.324 &  F & F\\
NGC5475 &     0.86 &    0.697 &    0.568 & G &     0.965 &     0.711 &      0.759 &     0.638 &  F & F\\
NGC5481 &     0.79 &    0.214 &    0.161 & G &     0.159 &     0.165 &      0.103 &     0.091 &  S & S\\
NGC5485 &     0.61 &    0.171 &    0.208 & G &     0.176 &     0.160 &      0.165 &     0.149 &  F & F\\
NGC5493 &     1.08 &    0.561 &    0.640 & G &     1.095 &     1.089 &      0.740 &     0.773 &  F & F\\
NGC5500 &     1.10 &    0.234 &    0.214 & G &     0.178 &     0.155 &      0.172 &     0.146 &  F & F\\
NGC5507 &     1.36 &    0.252 &    0.144 & G &     0.479 &     0.302 &      0.495 &     0.306 &  F & F\\
NGC5557 &     0.60 &    0.169 &    0.157 & G &     0.048 &     0.044 &      0.049 &     0.045 &  S & S\\
NGC5574 &     1.04 &    0.627 &    0.566 & G &     0.443 &     0.371 &      0.452 &     0.382 &  F & F\\
NGC5576 &     0.76 &    0.306 &    0.310 & G &     0.097 &     0.088 &      0.102 &     0.091 &  S & S\\
NGC5582 &     0.63 &    0.320 &    0.321 & G &     0.676 &     0.668 &      0.564 &     0.567 &  F & F\\
NGC5611 &     1.57 &    0.559 &    0.485 & G &     0.816 &     0.661 &      0.691 &     0.590 &  F & F\\
NGC5631 &     0.81 &    0.127 &    0.167 & G &     0.171 &     0.193 &      0.110 &     0.166 &  S & F\\
NGC5638 &     0.61 &    0.091 &    0.079 & G &     0.253 &     0.222 &      0.265 &     0.229 &  F & F\\
NGC5687 &     0.76 &    0.379 &    0.349 & G &     0.486 &     0.440 &      0.479 &     0.435 &  F & F\\
NGC5770 &     0.99 &    0.061 &    0.061 & G &     0.242 &     0.171 &      0.273 &     0.174 &  F & F\\
NGC5813 &     0.38 &    0.170 &    0.170 & G &     0.161 &     0.161 &      0.071 &     0.071 &  S & S\\
NGC5831 &     0.81 &    0.136 &    0.203 & G &     0.088 &     0.093 &      0.063 &     0.065 &  S & S\\
NGC5838 &     0.67 &    0.361 &    0.297 & G &     0.521 &     0.464 &      0.521 &     0.460 &  F & F\\
NGC5839 &     1.03 &    0.100 &    0.169 & G &     0.385 &     0.269 &      0.430 &     0.298 &  F & F\\
NGC5845 &     2.57 &    0.264 &    0.236 & G &     0.402 &     0.368 &      0.404 &     0.358 &  F & F\\
NGC5846 &     0.38 &    0.062 &    0.062 & R &     0.037 &     0.037 &      0.032 &     0.032 &  S & S\\
NGC5854 &     0.79 &    0.575 &    0.426 & G &     0.761 &     0.531 &      0.678 &     0.515 &  F & F\\
NGC5864 &     0.59 &    0.695 &    0.658 & G &     0.630 &     0.556 &      0.603 &     0.550 &  F & F\\
NGC5866 &     0.44 &    0.566 &    0.566 & G &     0.349 &     0.349 &      0.319 &     0.319 &  F & F\\
NGC5869 &     0.85 &    0.245 &    0.244 & G &     0.433 &     0.387 &      0.428 &     0.383 &  F & F\\
NGC6010 &     0.84 &    0.742 &    0.539 & G &     0.703 &     0.555 &      0.679 &     0.556 &  F & F\\
NGC6014 &     0.58 &    0.419 &    0.434 & G &     0.418 &     0.378 &      0.387 &     0.353 &  F & F\\
NGC6017 &     2.53 &    0.455 &    0.405 & G &     0.441 &     0.352 &      0.429 &     0.340 &  F & F\\
NGC6149 &     1.65 &    0.325 &    0.279 & G &     0.656 &     0.579 &      0.563 &     0.513 &  F & F\\
NGC6278 &     1.02 &    0.409 &    0.401 & G &     0.521 &     0.368 &      0.576 &     0.411 &  F & F\\
NGC6547 &     1.22 &    0.674 &    0.456 & G &     0.616 &     0.410 &      0.632 &     0.444 &  F & F\\
NGC6548 &     0.66 &    0.107 &    0.107 & G &     0.315 &     0.254 &      0.326 &     0.261 &  F & F\\
NGC6703 &     0.68 &    0.019 &    0.017 & G &     0.043 &     0.038 &      0.041 &     0.035 &  S & F\\
NGC6798 &     0.95 &    0.461 &    0.372 & G &     0.488 &     0.322 &      0.483 &     0.310 &  F & F\\
NGC7280 &     0.82 &    0.363 &    0.377 & G &     0.606 &     0.538 &      0.557 &     0.503 &  F & F\\
NGC7332 &     0.85 &    0.674 &    0.469 & G &     0.490 &     0.291 &      0.561 &     0.338 &  F & F\\
NGC7454 &     0.65 &    0.364 &    0.363 & R &     0.106 &     0.091 &      0.094 &     0.067 &  S & S\\
NGC7457 &     0.62 &    0.470 &    0.438 & G &     0.546 &     0.478 &      0.519 &     0.465 &  F & F\\
NGC7465 &     2.17 &    0.364 &    0.406 & G &     0.343 &     0.352 &      0.283 &     0.294 &  F & F\\
NGC7693 &     1.34 &    0.233 &    0.275 & G &     0.686 &     0.440 &      0.616 &     0.408 &  F & F\\
NGC7710 &     1.84 &    0.581 &    0.548 & S &     0.293 &     0.136 &      0.340 &     0.137 &  F & S\\
PGC016060 &     1.32 &    0.694 &    0.681 & G &     1.019 &     0.682 &      0.759 &     0.616 &  F & F\\
PGC028887 &     1.49 &    0.323 &    0.312 & G &     0.282 &     0.332 &      0.145 &     0.252 &  S & F\\
PGC029321 &     2.32 &    0.140 &    0.146 & G &     0.339 &     0.187 &      0.336 &     0.175 &  F & F\\
PGC035754 &     2.60 &    0.275 &    0.333 & G &     0.258 &     0.275 &      0.210 &     0.266 &  F & F\\
PGC042549 &     1.57 &    0.369 &    0.450 & G &     0.771 &     0.560 &      0.673 &     0.529 &  F & F\\
PGC044433 &     3.00 &    0.335 &    0.184 & S &     0.362 &     0.205 &      0.357 &     0.200 &  F & F\\
PGC050395 &     1.58 &    0.233 &    0.240 & G &     0.149 &     0.111 &      0.138 &     0.089 &  S & S\\
PGC051753 &     1.54 &    0.549 &    0.538 & G &     0.668 &     0.538 &      0.587 &     0.513 &  F & F\\
PGC054452 &     1.17 &    0.189 &    0.175 & G &     0.435 &     0.316 &      0.416 &     0.307 &  F & F\\
PGC056772 &     1.87 &    0.493 &    0.467 & G &     0.386 &     0.443 &      0.310 &     0.420 &  F & F\\
PGC058114 &     1.95 &    0.185 &    0.213 & R &     0.228 &     0.159 &      0.176 &     0.132 &  F & F\\
PGC061468 &     1.58 &    0.231 &    0.230 & G &     0.464 &     0.391 &      0.409 &     0.360 &  F & F\\
PGC071531 &     2.35 &    0.305 &    0.235 & R &     0.323 &     0.263 &      0.333 &     0.266 &  F & F\\
PGC170172 &     2.08 &    0.195 &    0.372 & G &     0.287 &     0.192 &      0.324 &     0.218 &  F & F\\
UGC03960 &     0.93 &    0.190 &    0.081 & G &     0.130 &     0.109 &      0.119 &     0.094 &  S & F\\
UGC04551 &     1.59 &    0.137 &    0.137 & G &     0.252 &     0.205 &      0.277 &     0.214 &  F & F\\
UGC05408 &     2.68 &    0.168 &    0.248 & G &     0.290 &     0.270 &      0.278 &     0.292 &  F & F\\
UGC06062 &     1.55 &    0.449 &    0.449 & G &     0.472 &     0.325 &      0.475 &     0.332 &  F & F\\
UGC06176 &     1.58 &    0.252 &    0.330 & G &     0.670 &     0.447 &      0.585 &     0.426 &  F & F\\
UGC08876 &     1.93 &    0.408 &    0.180 & S &     0.311 &     0.182 &      0.341 &     0.189 &  F & F\\
UGC09519 &     2.48 &    0.484 &    0.407 & G &     0.764 &     0.631 &      0.632 &     0.552 &  F & F\\
\hline
  \enddata 
  \tablecomments{Columns (1): Galaxy Name from the principal designation from LEDA; (2): largest equivalent aperture radius reached by the \Sauron\ maps in units of $R_e$; (3) and (4): moment ellipticity measured within one effective radius $R_e$ and one half effective radius $R_e/2$, these being replaced by the global ellipticity measured from the outer isophotes for galaxies with clear bars (see text) as indicated by a "C" in column (5) of this table ; (5) "Band" refers to the source of the radial ellipticity profiles, with G, R and S referring to the Green, Red and Sauron photometry, and "C" for galaxies which have strong bars significantly affecting the measurement of $\epsilon_e$, and for which the outer ellipticity measurement $\epsilon_{glob}$ (see Paper~II) is used instead; (6) and (7) $V/\sigma$ as measured within 1~$R_e$ and $R_e/2$; (8) and (9): $\lambda_R$ measured within 1~$R_e$ and $R_e/2$; (10) and (11) Fast (F) or Slow (S) Rotator according the to the $\lambda_R$ and $\epsilon$ values at $R_e$ and $R_e/2$. The present table (B1) is also available from our project website http://purl.com/atlas3d.} 
  \label{tab:LRparam} 
  \end{deluxetable}

\end{document}